\documentclass[a4paper,twoside,prd,preprintnumbers,twocolumn]{revtex4}
\usepackage{amssymb,amsmath,bm,natbib}
\usepackage{color}
\usepackage{slashed}
\usepackage{graphics}
\usepackage{graphicx}
\usepackage[utf8]{inputenc}
\usepackage{tabularx}
\usepackage[caption=false]{subfig}

\usepackage{hyperref}
\usepackage{url}

\PassOptionsToPackage{hyphens}{url}\usepackage{hyperref}

\usepackage{dsfont}
\usepackage{float} 
\usepackage{cancel}
\usepackage{ulem}

\begin{document}
	\preprint{PSI-PR-23-46, ZU-TH 77/23}
\title{\boldmath Light New Physics in $B\to K^{(*)}\nu\bar\nu$?}
	
	\author{Wolfgang Altmannshofer}
	\email{waltmann@ucsc.edu}
	\affiliation{Santa Cruz Institute for Particle Physics and Department of Physics, University of California, Santa Cruz, CA 95064, USA}

\author{Andreas Crivellin}
	\email{andreas.crivellin@cern.ch}
 		\affiliation{Physik-Institut, Universit\"at Z\"urich, Winterthurerstrasse 190, CH--8057 Z\"urich, Switzerland}
   \affiliation{Paul Scherrer Institut, CH--5232 Villigen PSI, Switzerland}

\author{Huw Haigh}
	\email{huw.haigh@oeaw.ac.at}
	\affiliation{Institute of High Energy Physics, 1050, Vienna, Austria}	
 
\author{Gianluca Inguglia}
	\email{gianluca.inguglia@oeaw.ac.at}
	\affiliation{Institute of High Energy Physics, 1050, Vienna, Austria}
	\author{Jorge Martin Camalich}
\email{jcamalich@iac.es}
\affiliation{Instituto de Astrof\'{\i}sica de Canarias, C/ V\'{\i}a L\'actea, s/n E38205 - La Laguna, Tenerife, Spain}
\affiliation{Universidad de La Laguna, Departamento de Astrof\'{\i}sica, La Laguna, Tenerife, Spain}
	
\begin{abstract}
The study of the rare decays $B\to K^{(*)} \nu \bar\nu$ offers a window into the dynamics operating at the electroweak scale, allowing studies of the Standard Model and searches for heavy new physics. However, the analysis of these decays is also potentially sensitive to the on-shell production of new light bosons $X$ through the process $B\to K^{(*)} X$. In particular, Belle~II has recently measured $B^+\to K^+\nu\bar\nu$, finding a $2.8\sigma$ excess under the assumption of heavy new physics. Since this excess is rather localized in the kaon energy, a fit that includes the decay mode $B^+\to K^+ X$ to the kinematic distributions prefers $m_X\approx2$\,GeV with branching fraction Br$[B\to KX]=(8.8\pm2.5)\times 10^{-6}$ and a significance of $\approx3.6\sigma$. However, no excess was found in the BaBar measurements of $B\to K^{(*)} \nu \bar\nu$, and a global analysis of the Belle II and BaBar data leads to Br$[B\to KX]=(5.1\pm2.1)\times 10^{-6}$ with a reduced significance of $\approx2.4\sigma$. We then study various simplified dark-flavoured models and present a possible UV completion based on a gauged $B_3-L_3$ symmetry, highlighting the discovery potential of dedicated searches for $B\to K^{(*)}X$ at Belle II.   
\end{abstract}
\maketitle

\section{Introduction}
\label{intro}

The Cabibbo-Kobayashi-Maskawa (CKM) mechanism~\cite{Kobayashi:1973fv} of the Standard Model (SM) was established by the $B$ factories Belle~\cite{Belle:2000cnh} and BaBar~\cite{BaBar:2001yhh} to be the leading source of quark flavour violation. Furthermore, the discovery of the Higgs boson~\cite{Higgs:1964ia,Englert:1964et} at the Large Hadron Colider (LHC) at CERN~\cite{ATLAS:2012yve,CMS:2012qbp} completed the SM. However, this does not exclude the existence of beyond-the-SM physics but rather only limits its possible size and strongly motivates the experimental search for it, both at the high-energy frontier and with precision observables.

Historically, indirect evidence for new particles often preceded direct discoveries. In particular, the existence of the charm quark, the $W$ boson, the top quark, and also the Higgs were expected due to indirect measurements of the Fermi interactions, kaon mixing, electroweak precision observables, etc. In this context, semi-leptonic $B$ meson decays are a particularly useful tool for indirect new physics (NP) searches: they have distinct and clean experimental signatures and, in general, controllable theoretical uncertainties as well as suppressed rates, making them sensitive probes of beyond-the-SM physics. In fact, an interesting number of anomalies, i.e.~deviations from the SM predictions arose~\cite{Crivellin:2023zui}. In particular, global fits to semileptonic $B$ decays involving $b\to c \tau\nu$ and $b\to s \ell^+\ell^-$ transitions show interesting hints for NP (see Refs.~\cite{Albrecht:2021tul,London:2021lfn,Capdevila:2023yhq} for recent reviews).

Recently, the Belle~II collaboration released an analysis of the closely related flavour-changing-neutral-current (FCNC) process $B^+\to K^+\nu\bar\nu$~\cite{Altmannshofer:2009ma, Buras:2014fpa} finding an excess of $2.8\sigma$ over the SM hypothesis.
This significance was obtained under the assumption of heavy NP~\cite{BptoKpnunuEPS}, allowing a connection to the anomalies in $b\to c \tau\nu$ and $b\to s \ell^+\ell^-$~\cite{Allwicher:2023syp,Athron:2023hmz,Felkl:2023ayn,Chen:2023wpb,He:2023bnk}. However, we will pursue a different path here: The $B\to K^+\nu\bar\nu$ measurement can be reinterpreted as a search for the two-body $B\to KX$ decay if the undetected particle $X$ is stable (approximately) or decays invisibly~\cite{MartinCamalich:2020dfe,Guerrera:2022ykl,Datta:2022zng,Bruggisser:2023npd,Abdughani:2023dlr,Berezhnoy:2023rxx}.~\footnote{Ref.~\cite{Dreiner:2023cms} considered light NP. However, the interactions are still mediated by heavy particles, such that no peak in the $q^2$ spectrum occurs.} For this, $X$ must be quite light, with $m_X\leq m_{B}-m_{K}$, such that it would result in a resonant feature in the spectrum of the squared invariant mass of the di-neutrino system (denoted by $q^2$) of $B\to K\nu\bar\nu$. This is different from the case of the SM, or any heavy NP contribution, where all $q^2$ dependence arises from the form factors, phase space, and the experimental efficiency.~\footnote{A light $Z^\prime$ boson in $b\to s\ell^+\ell^-$ has been proposed and studied in the literature~\cite{Sala:2017ihs,Mohapatra:2021izl,Datta:2017ezo,Altmannshofer:2017bsz,Sala:2018ukk,Bishara:2017pje,Borah:2020swo,Darme:2021qzw,Greljo:2021npi}, however it cannot explain the $b\to s\ell^+\ell^-$ anomalies because of di-muon invariant mass distribution in Drell-Yan production close to the $Z$ mass~\cite{Bishara:2017pje} or $e^+e^-\to \mu^+\mu^-+$invisible at Belle~II~\cite{Adachi:2019otg} and $B\to K^{(*)}\nu\bar\nu$~\cite{Belle:2017oht}.} 

Actually, the study of these types of experimental signatures has recently intensified in the context of dark-flavoured sectors, which are new light particles weakly coupled to the SM fermions with a rich flavour structure that can induce FCNC (see \cite{Kamenik:2011vy,Goudzovski:2022vbt} for reviews). This includes a QCD axion emerging from the breaking of horizontal flavour symmetries~\cite{Wilczek:1982rv,Feng:1997tn,Calibbi:2016hwq,Ema:2016ops,Arias-Aragon:2017eww,Choi:2017gpf,MartinCamalich:2020dfe,Calibbi:2020jvd,DiLuzio:2023ndz}, axion-like particles (ALPs)~\cite{Gripaios:2009pe,Izaguirre:2016dfi,Dolan:2017osp,Bjorkeroth:2018dzu,Gavela:2019wzg,Carmona:2021seb,Bauer:2021mvw,Ferber:2022rsf} and new neutral gauge bosons such as light $Z^\prime$ bosons in the closely related process $b\to s\ell^+\ell^-$~\cite{Sala:2017ihs,Mohapatra:2021izl,Datta:2017ezo,Altmannshofer:2017bsz,Sala:2018ukk,Bishara:2017pje,Borah:2020swo,Darme:2021qzw,Greljo:2021npi,Crivellin:2022obd}. 

Fortunately, Belle II provides information on the $q^2$ spectrum in their analysis, and, in fact, it shows a peak, localized around $q^2 = 4$\,GeV$^2$, suggesting that the possible excess might be better described by a new light mediator than with heavy NP. Therefore, in the next section, we will consider the experimental status of the $b\to s\nu\bar\nu$ transitions, including previous data from BaBar~\cite{BaBar:2013npw}, and perform a global analysis and recast of the data under the hypothesis of light NP. In Sec.~\ref{sec:models}, we will then study a series of simplified models that could be searched for by dedicated analyses of $b\to s X$ and propose an example of a possible UV completion before we conclude in Sec.~\ref{conclusion}.

\begin{figure*}[t]
     \centering
        \includegraphics[scale=0.4]{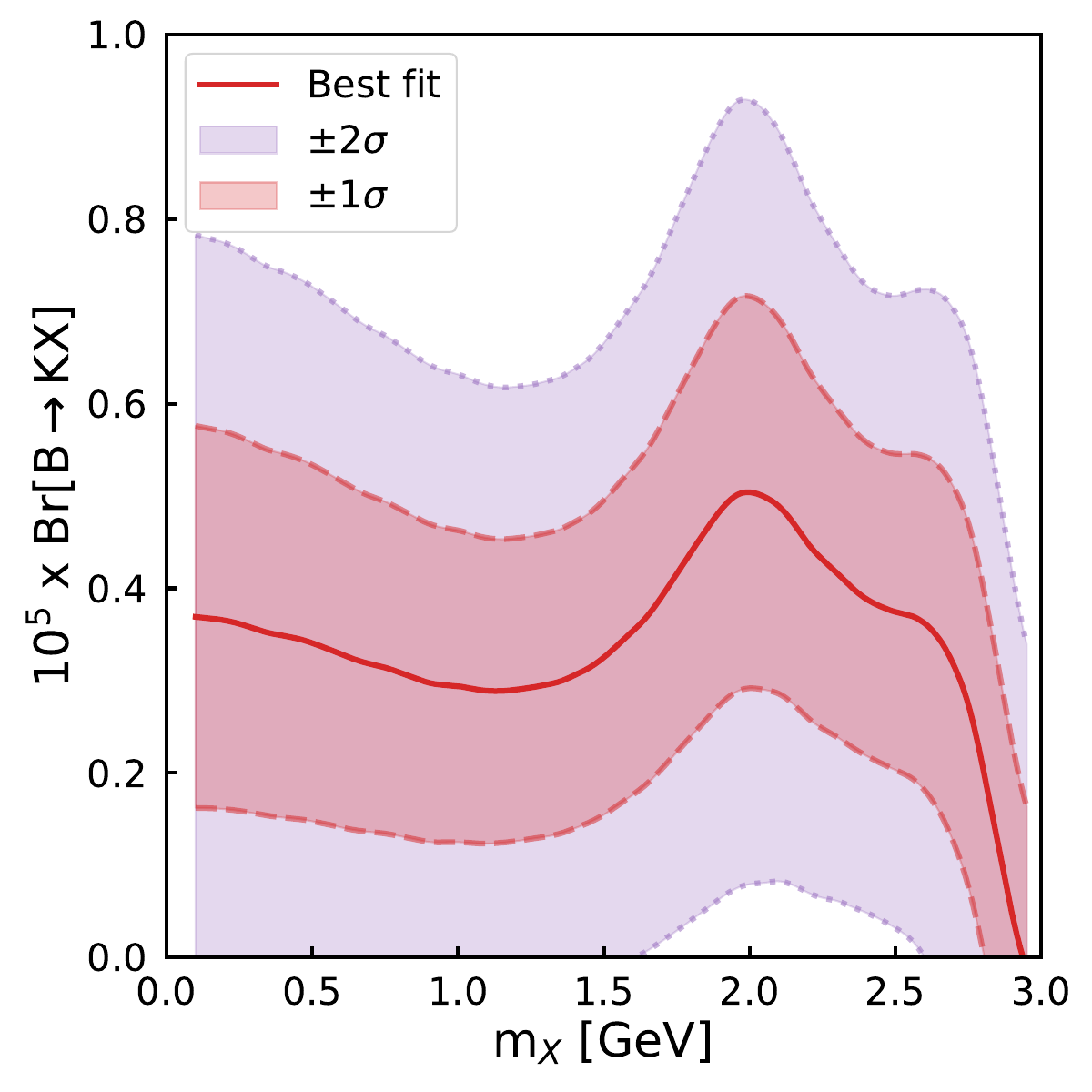}
        \includegraphics[scale=0.4]{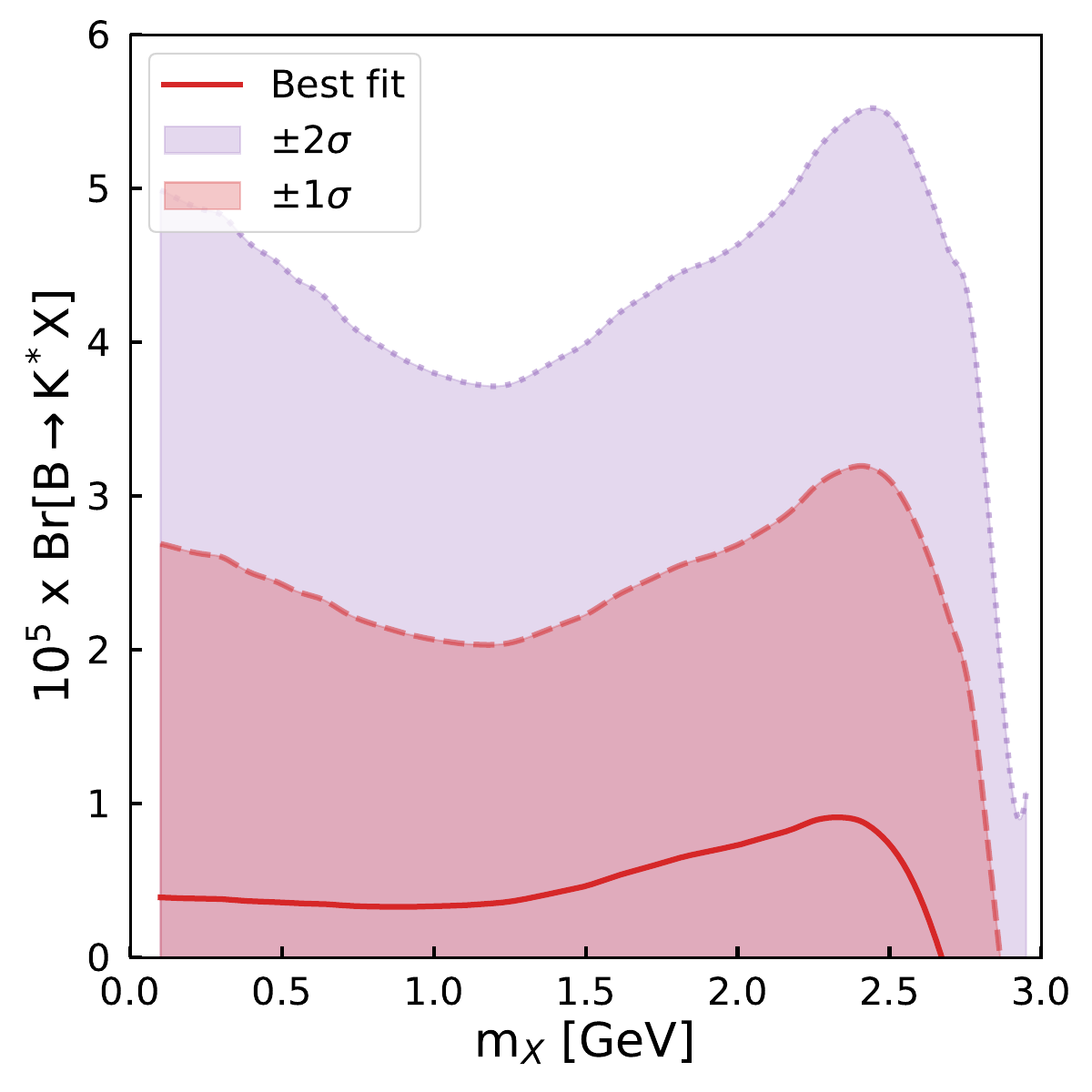}
        \caption{Left: Combined fit to Br$[B \to K X]$ from Belle II and BaBar as a function of the mass of $X$. Right: Same for Br$[B \to K^{*} X]$ (only BaBar data available).}
        \label{fig:bestFits}
\end{figure*}

\section{Experimental status and statistical analysis}
\label{sec:exp}

Due to the challenges in reconstructing the events in which the visible final state only involves a kaon (for charged $B$ meson decays) or its decay products (for neutral $B$ meson decays), searches for $B \to K^{(*)}\nu\bar\nu$ have only been performed at the $B$-factories Belle and BaBar.\footnote{Also a CLEO~\cite{CLEO:2000yzg} analysis exists which is however not competitive.} Here, $B$ mesons are produced in pairs from the decay of a $\Upsilon(4S)$ resonance and various analysis techniques to search for $B \to K^{(*)}\nu\bar\nu$ are available. In semileptonic tagged analyses (STA), one of the $B$ mesons is reconstructed via its semileptonic decay while the other one is used to search for $B\to K^{(*)}\nu\bar\nu$. Similarly, hadronic-tagged analyses (HTA) use the hadronic decay of one of the $B$ mesons and study the decay of interest of the other one. BaBar and Belle have searched for $B \to K^{(*)}\nu\bar\nu$ using both techniques~\cite{BaBar:2010oqg,BaBar:2013npw,Belle:2013tnz,Belle:2017oht}. The BaBar experiment found Br[$B^+ \to K^+\nu\bar\nu$]=$1.5^{+1.7+0.4}_{-0.8-0.2}\times10^{-5}$ and Br[$B \to K^*\nu\bar\nu$]=$3.8^{+2.9}_{-2.6}\times10^{-5}$~\cite{BaBar:2013npw}, while Belle provided 90\% CL upper limits to the same process at the level of $2.7\times10^{-5}$.

An additional analysis technique, referred to as inclusive tag analysis (ITA), already adopted by Belle and Belle~II in previous studies~\cite{Belle:2019iji,Belle-II:2021rof}, allows one to reconstruct \textit{inclusively} the decay $B^+ \to K^+\nu\bar\nu$ from the charged kaon. This alternative methodology negates the requirement of a coincidental fully reconstructed hadronic (or semileptonic) $B$-decay to tag the event, thereby providing a higher signal efficiency at the cost of reduced signal purity due to increased background levels. In the recently released results obtained by Belle~II, both ITA and HTA techniques are used~\cite{BptoKpnunuEPS}, and the results are combined. Driven by the ITA technique with its higher statistics, Belle~II obtained the first evidence for the decay $B^+ \to K^+\nu\bar\nu$ with 3.6$\sigma$ significance, measuring Br[$B^+ \to K^+\nu\bar\nu$]=$(2.4\pm0.7)\times10^{-5}$. This is in 2.8$\sigma$ tension with the SM prediction of Br$[B^+ \to K^+\nu\bar\nu]_{\rm SM}=(0.497\pm0.037) \times10^{-5}$ (excluding the contribution from $B^+ \to \tau^+\nu$ with $\tau^+ \to K^+\bar\nu$)~\cite{Parrott:2022zte} (see also~\cite{Becirevic:2023aov,Amhis:2023mpj} for other recent SM predictions that agree well with the result that we are using).\footnote{If one considers specifically the Belle II ITA, which is the most sensitive of the two presented, the significance of the signal is slightly higher, providing Br[$B^+ \to K^+\nu\bar\nu$]=$(2.8\pm0.7)\times10^{-5}$, which is in tension with the SM predictions at the 3.3 $\sigma$ level. We assume that the two analyses, ITA and HTA, select two independent populations of events; in other words, we assume no statistical correlation between the two individual ITA and HTA measurements, which is justified by the small sample overlap.} 

This result is interesting because it might indicate not only the presence of NP in the $b\to s \nu\bar\nu$ transitions but even the presence of new light states. This can be seen by looking at the supplemental material that accompanies the Belle~II result~\cite{BptoKpnunuEPS}. The post-fit distributions of events as a function of $q^2$ indicate that the observed excess clusters in the region around $q^2 = 4$\,GeV$^2$, as can be seen in the right plot of Fig.~\ref{fig:signal_region_histograms} in the appendix, showing the data and SM yields from the Belle~II search for $B^+\to K^+\nu \bar\nu$~\cite{BptoKpnunuEPS}. However, to evaluate the significance of such an excess, a fit taking into account the experimental resolution and all available data, including BaBar's where no excess has been observed, has to be performed. 

Therefore, we use the differential distributions of the $B^{0,+}\to K^{0,+}\nu\bar\nu$ measurements of Belle~II~\cite{BptoKpnunuEPS} and BaBar~\cite{BaBar:2013npw} under the assumption that a light resonance escaping detection is present (i.e.~$X$ is either stable, sufficiently long-lived or decays invisibly) to evaluate the combined significance for NP. Furthermore, we will use the $B\to K^{*}\nu\bar\nu$ measurement of BaBar~\cite{BaBar:2013npw} to set an upper limit on Br$[B\to K^*X]$. 

We fit the NP signal to the reconstructed data by modelling the resonance $X$ with a Gaussian distribution. This is done via a binned maximum likelihood fit, using the \texttt{pyhf} software package~\cite{Heinrich:2021gyp}. In the combined fit to $B^+\to K^+\nu\bar\nu$ (Belle and Babar) and $B^0\to K^0\nu\bar\nu$ (BaBar) data, each measurement constitutes a channel in the statistical \texttt{pyhf} model with a fully correlated signal. Similarly, in the fit to the BaBar $B^0\to K^{0,*}\nu\bar\nu$ and $B^+\to K^{+,*}\nu\bar\nu$ distributions, the relative signal is fixed by isospin invariance to be (approximately) equal.  

While we assume that the particle $X$ has a negligible intrinsic (physical) width, we nonetheless assign a Gaussian standard deviation of 1.5$\,$GeV$^2$ to its $q^2$ distribution to capture the detector resolution.\footnote{We checked that the significance is to a good approximation independent of the function used to parameterize the resonance $X$ as long as its width captures the resolution of the detector. However, if e.g.~a Crystal-ball function is used instead, a slightly
lower mass of $X$ is obtained.}  

The fit to the Belle-II data alone results in  ${\rm Br}[B^+\to K^+X=(8.8\pm2.5)\times 10^{-6}$ for $m_X\approx2\,$GeV, with a significance of $3.6\sigma$. The inclusion of BaBar data in the fit reduces the significance to $2.4\sigma$ and 
 \begin{equation}\label{eq:BtoKXglobal}
     {\rm Br}[B\to KX]=(5.1\pm2.1)\times 10^{-6}\,,
 \end{equation}
is preferred (see Fig.~\ref{fig:BF_bestFits} in the appendix for details). For $B\to K^*X$, only BaBar data is available, and since there is no excess seen, an upper limit of a few times $10^{-5}$, depending on $m_X$, can be obtained. The results of the $B\to K X$ and $B\to K^{*}X$ fits are depicted in the left and right panels of Fig.~\ref{fig:bestFits}, respectively.

\section{Models of Light New Physics}
\label{sec:models}

Here, we consider two options of light particles that can lead to $B\to KX$, a light neutral vector (i.e.~a $Z^\prime$) and flavoured axions and ALPs~\cite{Wilczek:1982rv,Feng:1997tn,Calibbi:2016hwq,Ema:2016ops,Arias-Aragon:2017eww,Choi:2017gpf,MartinCamalich:2020dfe,Calibbi:2020jvd,DiLuzio:2023ndz,Gripaios:2009pe,Izaguirre:2016dfi,Dolan:2017osp,Bjorkeroth:2018dzu,Gavela:2019wzg,Carmona:2021seb,Bauer:2021mvw,Ferber:2022rsf}. In both cases, $X$ should not decay to charged SM fermions as those decays would give prominent resonant signals in e.g.~$b \to s \ell^+ \ell^-$ decays. Couplings to electrons, muons, and light quarks should be absent or sufficiently small such that the new boson is long-lived enough to decay outside the detector or has a dominant invisible decay width.~\footnote{This requirement excludes the often studied $L_\mu-L\tau$ $Z^\prime$ boson~\cite{He:1991qd, Altmannshofer:2014cfa}.} 

Since a flavour-changing bottom-strange coupling ($g_{sb}$) is needed to obtain the desired decay mode, in principle constraints from $B_s - \bar B_s$ mixing have to be considered. In fact, for a light $Z^\prime$ or ALP, one can set up an operator product expansion in $m_X/m_b$ to calculate this new physics contribution and obtain bounds on the flavour changing couplings $g_{sb}$~\cite{MartinCamalich:2020dfe}. However, these limits are typically weaker than the ones obtained from decays such as $B\to K^{(*)}X$ because (in contrast to the case of heavy NP) the decay rate is proportional to the quadratic (not quartic) power of the couplings which is the same scaling as for the neutral-meson mixing amplitude. 

\subsection{\boldmath Light vectors ($Z^\prime$)}

\begin{figure*}[t]
\centering 
\includegraphics[width=0.45\textwidth]{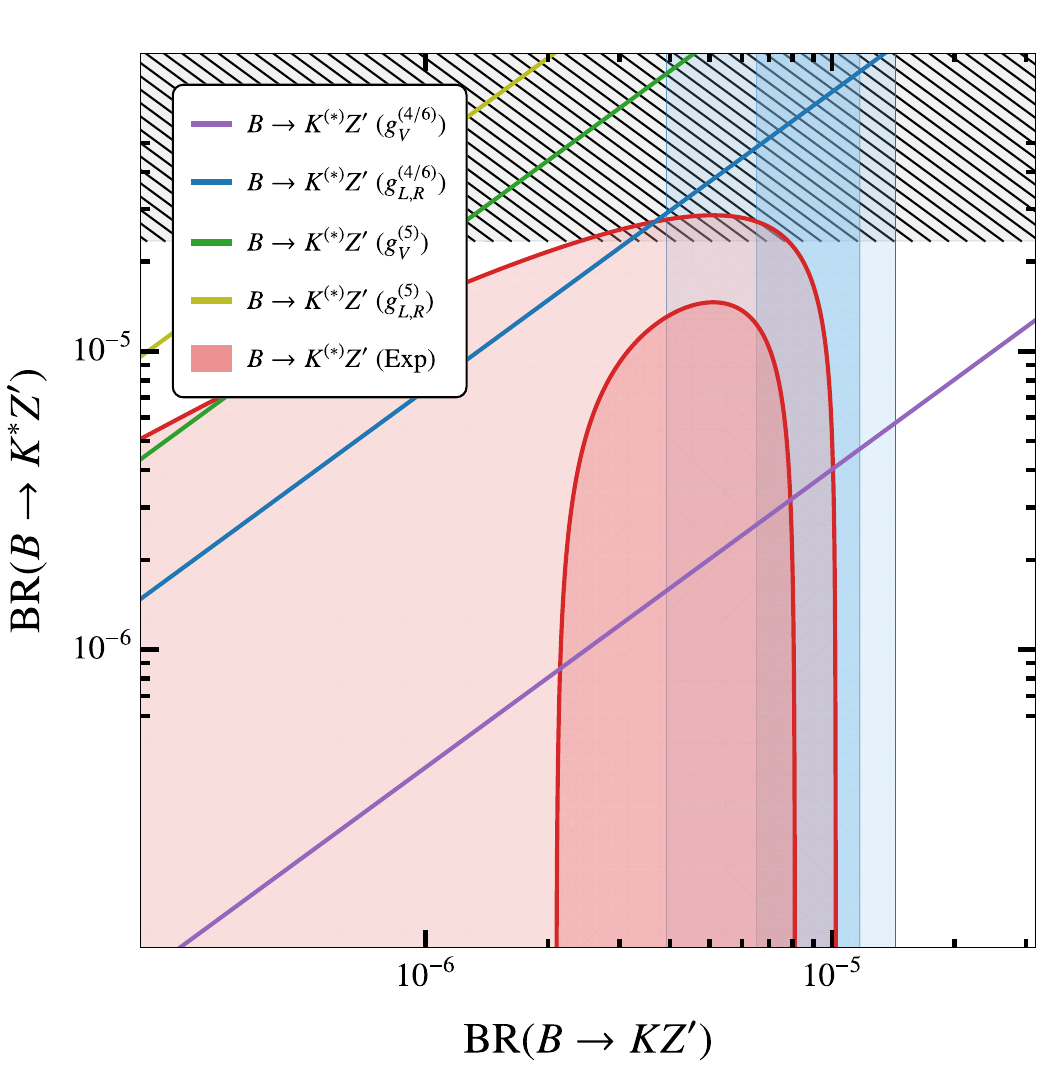}~~~
 \includegraphics[width=0.44\textwidth]{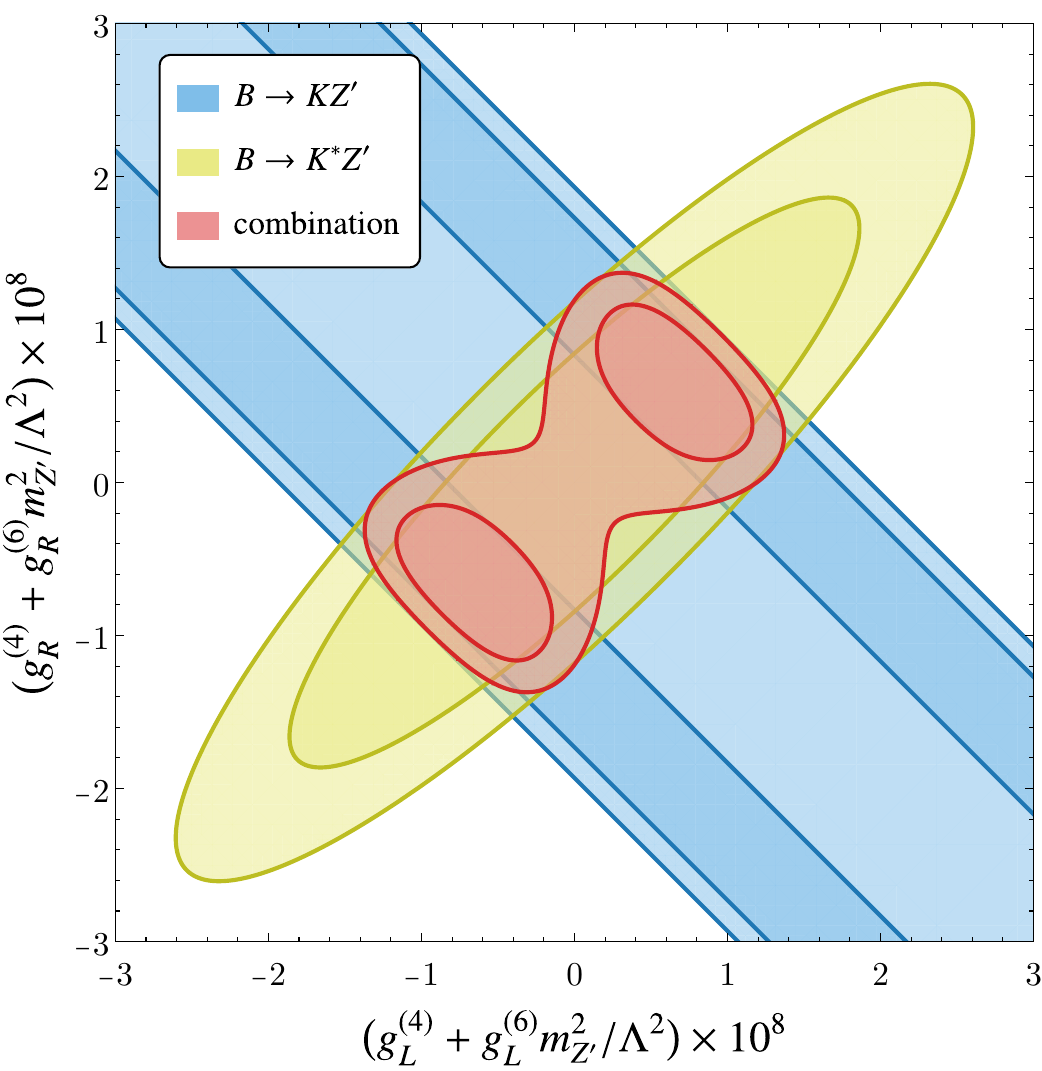}
	\caption{\textit{Left:} Correlations between $B\to KZ^\prime$ and $B\to K^*Z^\prime$ (colored lines) for the different $\bar s bZ^\prime$ operators considered in this work. These are compared to the experimental data stemming from the combination of Belle-II, Babar and Belle measurements,which is represented by the red regions correspondingto $\Delta \chi^2=2.3$ and $\Delta \chi^2=6.18$.
    Belle's upper limit (hatched region at 2$\sigma$) and the new Belle II measurement (blue vertical band at 1$\sigma$ and 2$\sigma$). \textit{Right:} preferred regions in the $g_L-g_R$ plane. One can see that (approximately) vectorial couplings of the order of $10^{-8}$ are suggested by current data.}
	\label{fig:Zprime}%
\end{figure*}

Including couplings up to dimension 6, the interaction Lagrangian is~\cite{Altmannshofer:2017bsz}
\begin{multline}
\mathcal L_{Z^\prime} \supset \Big\{ g_L^{(4)} Z^\prime_\mu (\bar s \gamma^\mu P_L b) + \frac{g_L^{(5)}}{\Lambda} Z^\prime_{\mu\nu} (\bar s \sigma^{\mu\nu} P_R b)\\
+ \frac{g_L^{(6)}}{\Lambda^2} \partial^\nu Z^\prime_{\mu\nu} (\bar s \gamma^\mu P_L b) ~+~ \text{h.c.} \Big\} ~+~ \{ L \leftrightarrow R \}\,,
\end{multline}
where $Z^\prime_{\mu \nu} = \partial_\mu Z^\prime_\nu -  \partial_\nu Z^\prime_\mu$ is the $Z^\prime$ field strength tensor. For later convenience, we also introduce vector and axial-vector couplings $g_V^{(d)} = g_R^{(d)} + g_L^{(d)}$ and $g_A^{(d)} = g_R^{(d)} - g_L^{(d)}$.

In this setup, we find the following $B \to K Z^\prime$ decay rates if only one of the couplings is switched on at a time
\begin{eqnarray}
\Gamma^{(4)}_{B \to K Z^\prime} &=& \frac{|g_V^{(4)}|^2}{64\pi} \frac{m_B^3}{m_{Z^\prime}^2} \lambda^\frac{3}{2} f_+ \,,\\
\Gamma^{(5)}_{B \to K Z^\prime} &=& \frac{|g_V^{(5)}|^2}{16\pi} \frac{m_B m_{Z^\prime}^2}{\Lambda^2}\left(1 + \frac{m_K}{m_B}\right)^{-2} \lambda^\frac{3}{2} f_T \,, \\
\Gamma^{(6)}_{B \to K Z^\prime} &=& \frac{|g_V^{(6)}|^2}{64\pi} \frac{m_B^3 m_{Z^\prime}^2}{\Lambda^4} \lambda^\frac{3}{2} f_+ \,,
\end{eqnarray}
with the phase space function
\begin{equation}
\lambda = 1 + \frac{m_K^4}{m_B^4} + \frac{m_{Z^\prime}^4}{m_B^4} - 2 \left( \frac{m_K^2}{m_B^2} + \frac{m_{Z^\prime}^2}{m_B^2} + \frac{m_K^2 m_{Z^\prime}^2}{m_B^4}\right)\,.
\end{equation}
The $B \to K$ form factors $f_+$ and $f_T$ can be found in Refs.~\cite{Parrott:2022rgu, Becirevic:2023aov, Gubernari:2023puw} and have to be evaluated at $q^2 = m_{Z^\prime}^2$. 

Similarly, we find for the $B \to K^\star Z^\prime$ decay rates
\begin{eqnarray}
\Gamma^{(4)}_{B \to K^* Z^\prime} &=& \frac{m_B}{32\pi} \lambda_*^\frac{1}{2} \Big[ |g_V^{(4)}|^2 \mathcal F_V + |g_A^{(4)}|^2 \mathcal F_A \Big] \,,\\
\Gamma^{(5)}_{B \to K^* Z^\prime} &=& \frac{m_B}{8\pi} \frac{m_B^2}{\Lambda^2} \lambda_*^\frac{1}{2} \Big[ |g_V^{(5)}|^2 \mathcal F_T + |g_A^{(5)}|^2 \mathcal F_{T5} \Big] \,, \\
\Gamma^{(6)}_{B \to K^* Z^\prime} &=& \frac{m_B}{32\pi} \frac{m_{Z^\prime}^4}{\Lambda^4} \lambda_*^\frac{1}{2} \Big[ |g_V^{(6)}|^2 \mathcal F_V + |g_A^{(6)}|^2 \mathcal F_A \Big] \,,
\end{eqnarray}
with
\begin{eqnarray}
\mathcal F_V &=& \lambda_* \left(1+\frac{m_{K^*}}{m_B}\right)^{-2} V^2 \,, \\
\mathcal F_A &=& \lambda_* \left(1+\frac{m_{K^*}}{m_B}\right)^2 A_1^2 + \frac{32 m_{K^*}^2}{m_{Z^\prime}^2} A_{12}^2 \,, \\
\mathcal F_T &=& \lambda_* T_1^2 \,, \\
\mathcal F_{T5} &=& \left(1-\frac{m_{K^*}^2}{m_B^2}\right)^2 T_2^2 + \frac{8 m_{K^*}^2 m_{Z^\prime}^2}{m_B^2 (m_B + m_{K^*})^2} T_{23}^2\,.
\end{eqnarray}
The phase space function $\lambda_*$ is defined equivalently to the $B \to K Z^\prime$ decay, with the replacement $m_K \to m_{K^*}$.
The $B \to K^*$ form factors $V$, $A_1$, $A_{12}$, $T_1$, $T_2$, $T_{23}$ can be found in Refs.~\cite{Bharucha:2015bzk,Gubernari:2023puw}.

The results can be seen in the plots of Fig.~\ref{fig:Zprime} that show the correlation between the $B \to K Z^\prime$ and $B \to K^* Z^\prime$ branching ratios (left plot) and the best-fit regions in the plane of $Z^\prime$ couplings to left-handed and right-handed quark currents (right plot). In both plots, the mass of the $Z^\prime$ is fixed to the best-fit value $\sim 2$\,GeV, c.f. the discussion in section~\ref{sec:exp}. One can see that couplings only to left-handed or right-handed quarks, $g_{L/R}^{(4/6)}$, generate $B \to K^* Z^\prime$ branching ratios that exceed the experimental bounds by a factor of few. The dimension-5 dipole couplings, $g_{L/R}^{(5)}$, lead to even larger $B \to K^* Z^\prime$. Couplings that are dominantly vectorial are needed for explaining the enhanced $B\to K\nu\bar\nu$ branching ratio without violating the constraints from $B\to K^*\nu\bar\nu$. The size of the $Z^\prime$ couplings is very small, at the order of $10^{-8}$. 

\begin{figure*}[t]
    \centering
        \includegraphics[scale=0.325]{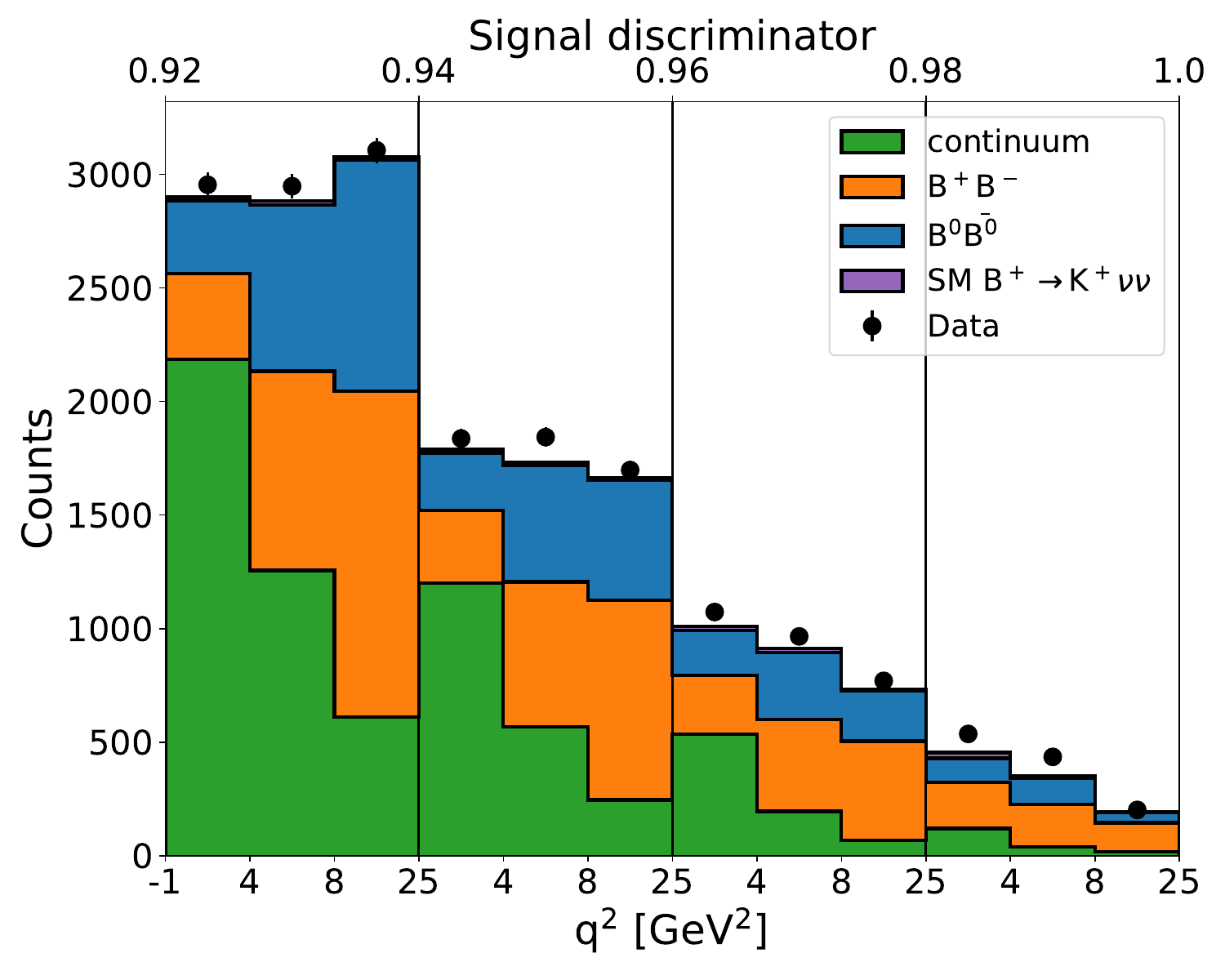}
        \includegraphics[scale=0.325]{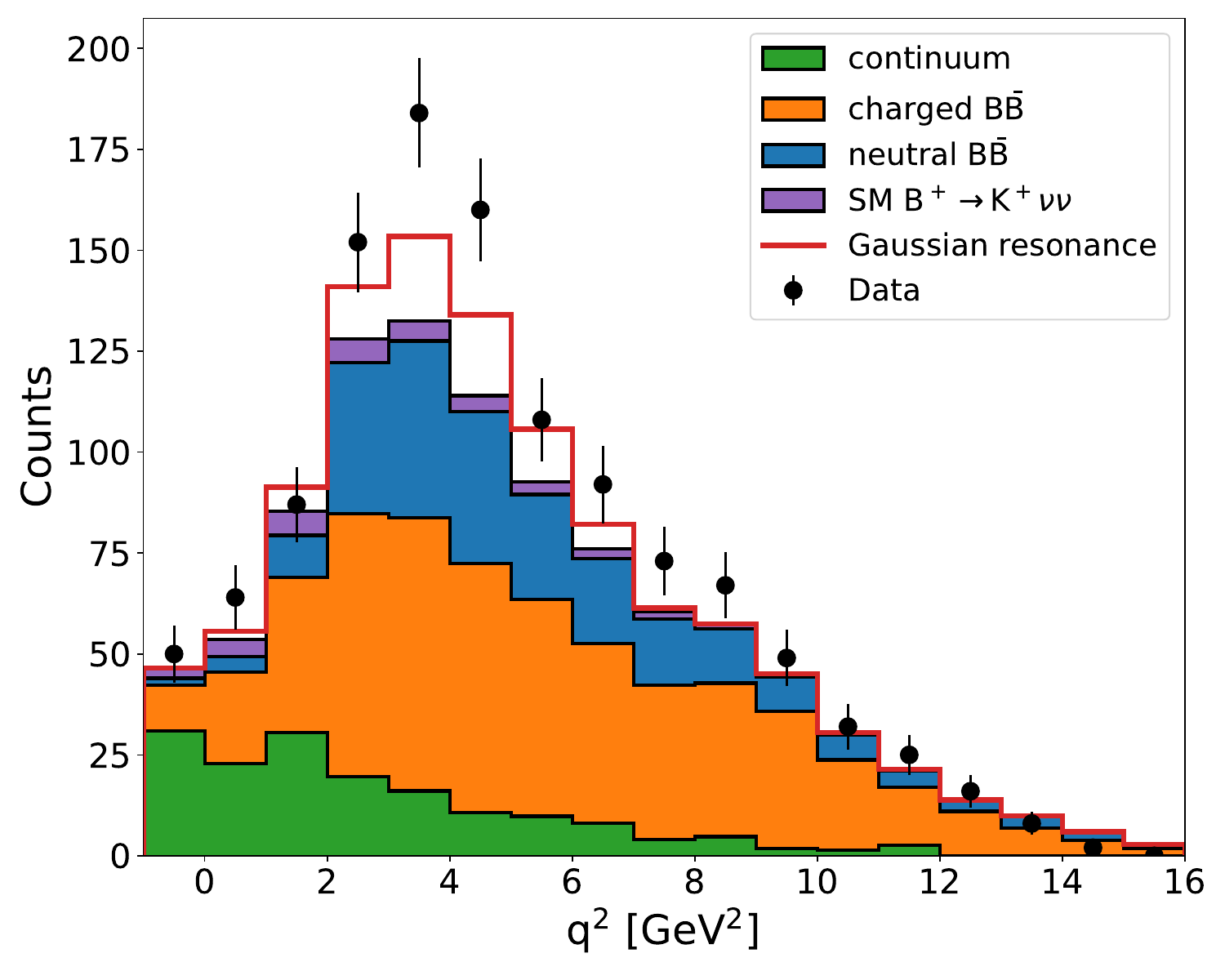}
        \caption{Left: Histogram showing Belle II data and MC~\cite{BptoKpnunuEPS} used in the NP fit, with four signal regions of the discriminator output, each containing 2\% signal efficiency across three $q^2$ bins. Right: The highest sensitivity bin (i.e.~with signal discriminator $0.98-1$) split into finer $q^2$ bins, showing the resonant characteristics of the observed excess for which the red line shows the best fit. Note that these data were not used for the fit but are only shown as illustrations.}
        \label{fig:signal_region_histograms}
\end{figure*}

\subsection{Axion like particle}

We consider now a massive pseudoscalar or axion-like particle (ALP) $a$ coupled with a derivative coupling to the $bs$ current,
\begin{align}
\mathcal L_{\rm ALP}\supset \frac{\partial_\mu a}{2f}\left(\bar s\gamma^\mu(g_{V}+g_{A}\gamma_5)b\right)+{\rm h.c.} 
\end{align}
where $f$ is the ALP decay constant and where we have started from the vectorial basis for the couplings. Note that this also covers the case of neutral (pseudo)scalars with (effective) $g_S\bar s  b+g_P\bar s \gamma_5 b$ couplings (without a derivative), by identifying $g_{V(A)} m_b/2f=g_{S(P)}$ as per the equations of motion. 

In this setup, the $B \to K a$ and $B \to K^* a$ decay rates are given by,
\begin{align} \label{eq:ALP_BK}
&\Gamma_{B\to K a}=\frac{|g_V|^2\,m_B^3}{64\pi f^2}\left(1-\frac{m_K^2}{m_B^2}\right)^2\lambda^\frac{1}{2}f_0^2 \,,\\  \label{eq:ALP_BKstar}
&\Gamma_{B\to K^* a}=\frac{|g_A|^2\,m_B^3}{64\pi f^2}\lambda_*^\frac{3}{2}A_0^2 \,,
\end{align}
where $\lambda$ and $\lambda^*$ are now the same as for $B\to K Z^\prime$ and $B\to K^* Z^\prime$, respectively, but replacing $m_{Z^\prime}$ by $m_a$. The form factors $f_0$ and $A_0$ did not enter the expressions for the $B\to K^{(*)} Z^\prime$ rates but can also be found in Refs.~\cite{Bharucha:2015bzk, Parrott:2022rgu, Becirevic:2023aov, Gubernari:2023puw}. They are evaluated at $q^2=m_a^2$. 

As is evident from equations~\eqref{eq:ALP_BK} and \eqref{eq:ALP_BKstar} above, only vectorial couplings are capable of explaining $B\to K\nu\bar\nu$ while axial vector couplings are constrained by $B\to K^*\nu\bar\nu$.
The corresponding constraints on the couplings are qualitatively similar to the $Z^\prime$. In terms of the ALP decay constant, Eq.~\eqref{eq:BtoKXglobal} implies $F_V\equiv2f/|g_V|=3.1^{+1.0}_{-0.5}\times10^8$ for $m_a=2$ GeV, while the upper limit from $B\to K^*a$ leads to $F_A\equiv2f/|g_A|\gtrsim1.7\times10^8$ GeV at 2$\sigma$.

\subsection{\boldmath $B_3-L_3$ Symmetry}

Let us outline one possible UV complete model that could give rise to $B\to K X$, which is based on a gauged $B_3-L_3$ symmetry. This means we assume that third-generation baryon and lepton numbers are oppositely charged under a new gauged $U(1)_X$ symmetry. This charge assignment is anomaly-free and can provide an explanation of the smallness of the CKM elements $V_{cb,ts}$ and $V_{ub,td}$. The reason for this is that in unbroken $U(1)_X$, no $1-3$ and $2-3$ elements in the quark Yukawa couplings are allowed. This means that while all quark masses and the Cabibbo angle can be obtained from the SM Higgs after it acquires its vacuum expectation value (VEV) from renormalizable  dim-4 couplings, $V_{cb,ts}$ and $V_{ub,td}$ are zero at this level. One option to obtain non-zero $V_{cb,ts}$ and $V_{ub,td}$ is to introduce additional Higgs doublets charged under $U(1)_X$ to generate $V_{cb,ts}$ and $V_{ub,td}$ via their VEVs (possibly in conjunction with vector-like quarks~\cite{Altmannshofer:2014cfa}). 

Since the $U(1)_X$ gauge boson needs to have both left-handed and right-handed $bs$ coupling to explain $B\to K X$ without violating the bounds from $B\to K^* X$, we have to add two additional Higgs doublets charged under $B_3-L_3$ with opposite $U(1)_X$ charges~\cite{Crivellin:2015lwa}.\footnote{Note that in case the new CP-odd and CP-even Higgses of each doublet are quasi-degenerate, the effect in $B_q-\bar B_q$ mixing is suppressed~\cite{Crivellin:2013wna}.}
Furthermore, a singlet under $SU(2)_L$ charged under $U(1)_X$ is needed to obtain the preferred $Z^\prime$ mass $m_{Z^\prime} \simeq 2$\,GeV without overshooting Br$[B\to KX]$. Finally, note that since 2\,GeV is below the bottom and tau thresholds, such a $Z^\prime$ boson naturally decays invisibly to tau neutrinos and thus escapes detection.

\begin{figure*}[t]
     \centering
        \includegraphics[scale=0.4]{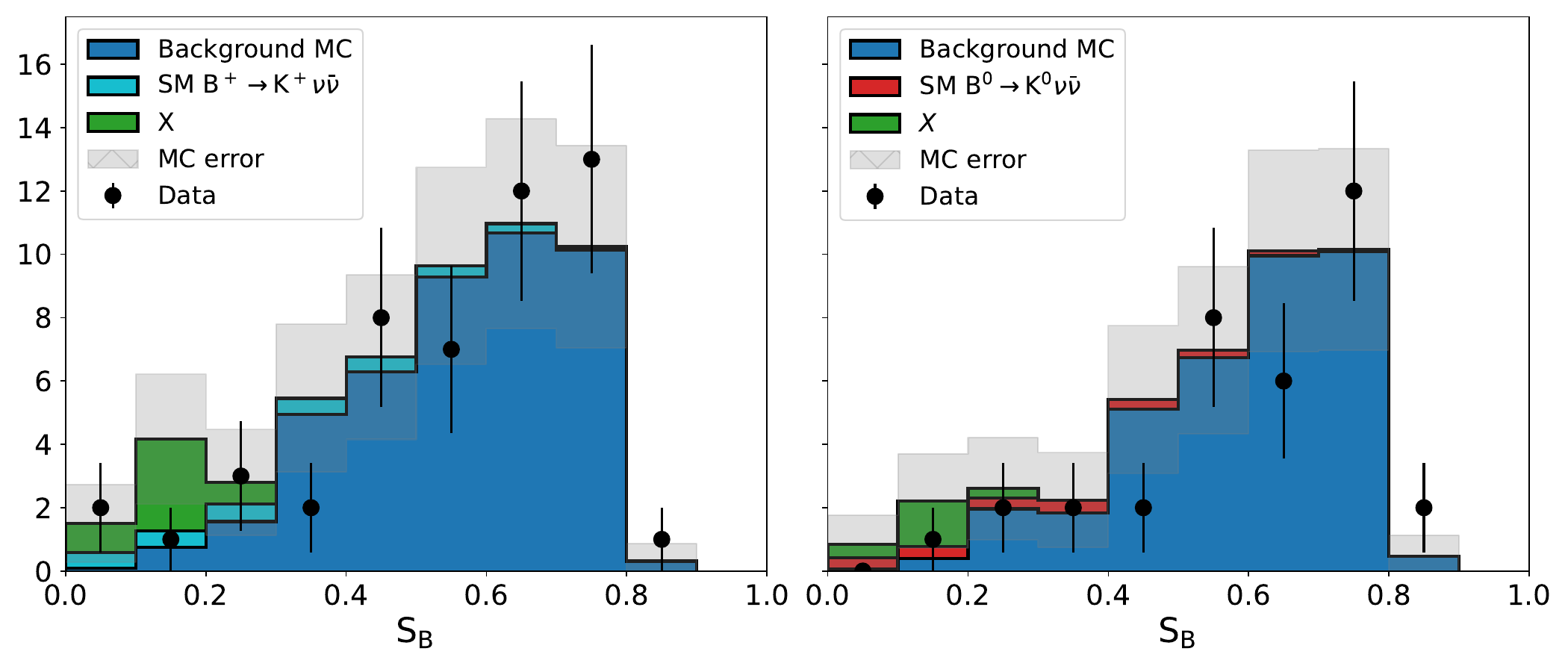}
        \caption{BaBar data and simulated MC~\cite{BaBar:2013npw}, showing $B^{+}\to K^{+}\nu\bar{\nu}$ (left) and $B^{0}\to K^{0}\nu\bar{\nu}$ (right), provided in bins of $S_B=q^2/m_B^2$. The distribution of the fitted resonance is shown in green.}
        \label{fig:babar_histograms}
\end{figure*}

\begin{figure}[t]
     \centering
        \includegraphics[scale=0.4]{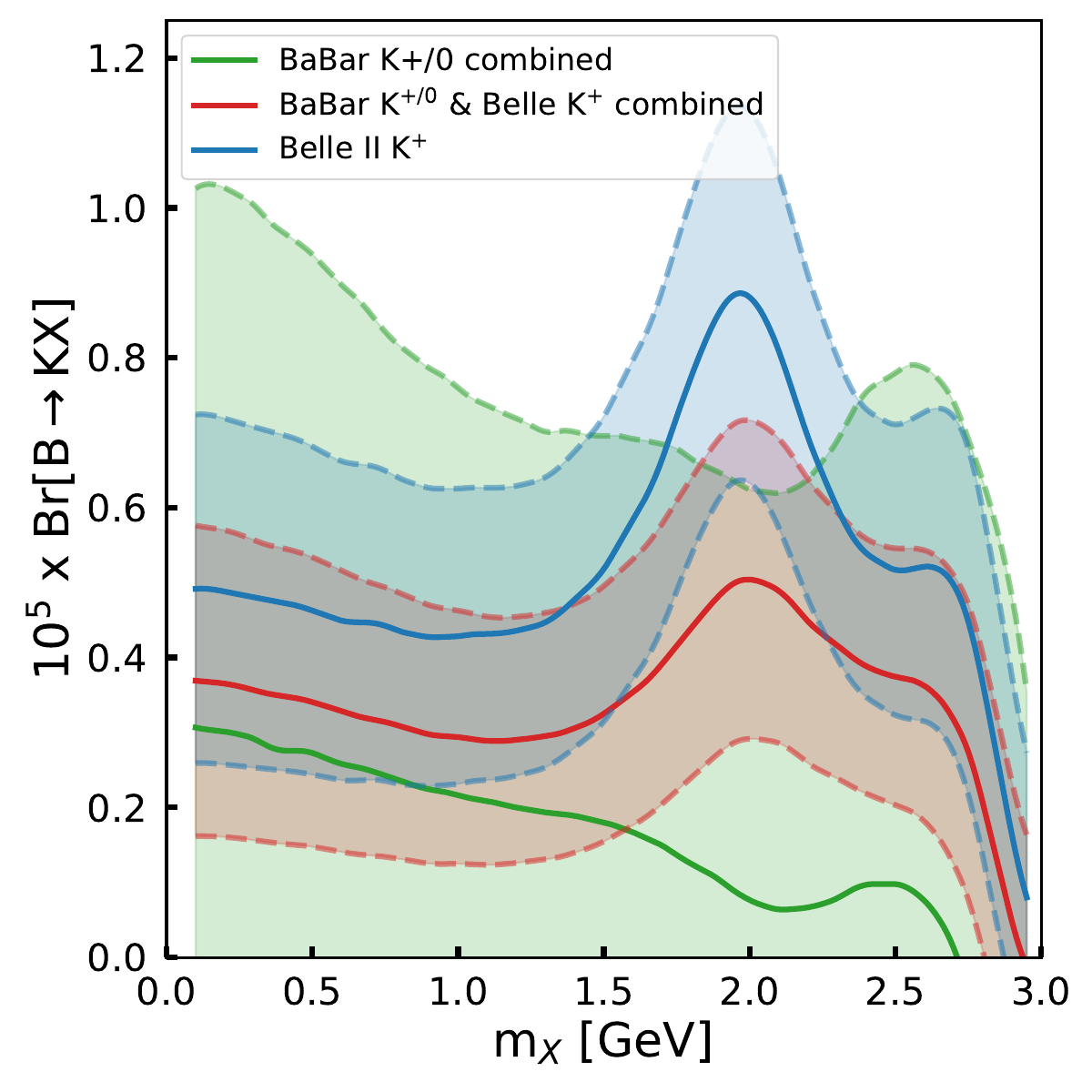}
        \caption{Best fit and associated $1\sigma$ errors for Br$[B\to K^{(*)}X]$ as a function of m$_X$, for the fit to the BaBar distributions (green), the Belle II distribution (blue) and the combined fit to all data (red).}
        \label{fig:BF_bestFits}
\end{figure}

\section{Conclusions and Outlook}
\label{conclusion}

Motivated by the recently observed excess of $2.8\sigma$ (w.r.t.~the SM prediction) in $B^+\to K^+\nu\bar\nu$ by Belle~II, we studied the option of light new physics, i.e.~$B\to K X$ where the new boson $X$ escapes detection. In fact, since the excess is localized around $4\,$GeV${}^2$, the investigation of light NP hypothesis is motivated. Assuming that the ``effective'' width of the particle is given by the detector resolution (i.e.~the physical width is small compared to the detector resolution), we found a significance of $\approx\!2.4\sigma$ and a preferred branching ratio of Br$[B\to KX] = (5.1 \pm 2.1)\times 10 ^{-6}$ for $m_X\approx 2\,$GeV once the Belle II data is combined with the corresponding BaBar result.\footnote{This differs from the results found in Ref.~\cite{He:2023bnk} where a mass of $X$ of a few hundred GeV was proposed.} Similarly we performed a fit to $B\to K^*\nu\bar\nu$ where no excess is observed.

We studied the two simplified NP physics models with a light vector and a light pseudoscalar with derivative couplings (i.e.~an ALP). It is found that the flavour-changing coupling to bottom and strange quarks should be dominantly vectorial to explain the excesses. Finally, we proposed an example of a UV complete model, a gauged $B_3-L_3$ symmetry, broken by three additional scalars (two $SU(2)_L$ doublets and one singlet) that naturally leads to the desired signature.

One should note that while our fitting approach provides insightful information on possible underlying processes in $B^+ \to K^+\nu\bar\nu$, it leaves room for improvements. The reason is that while the $B^+ \to K^+ X$ carries kinematic information typical of a two-body decay, the original search is not optimized for such a case, and a dedicated experimental analysis will provide a better sensitivity. However, we hope that our work motivates dedicated analyses by the $B$ factories.

\appendix

\section{Details on the fit}

We perform maximum likelihood fits to the BaBar~\cite{BaBar:2013npw} and Belle II~\cite{BptoKpnunuEPS} data using the \texttt{pyhf} software package~\cite{Heinrich:2021gyp}. For this, we include the essential estimated experimental systematic uncertainties, based on those quoted in the Belle II analysis, as nuisance parameters. The signal is fit to data using binned templates of the $q^2$ distributions, derived from post-fit Monte Carlo (MC) distributions, including individual contributions from $B\bar{B}$ and continuum SM background and the predictions of the SM contribution to $B\to K^{(*)}\nu\nu$. The corresponding templates are shown in Fig.~\ref{fig:signal_region_histograms} (left) for the Belle~II analysis. They contain 12 bins in total: Three $q^2$ bins which are repeated in four bins of the signal discriminator output. These are constructed such that the expected signal efficiency is a constant 2\% in the four regions. 

For each of the four fit templates shown on the left in Fig.~\ref{fig:signal_region_histograms}, we include an overall normalization uncertainty of 10\% and the associated statistical uncertainty obtained from the measured number of events. The continuum background template has an additional uncertainty of 10\% from shape systematics, which we allow for each bin to fluctuate individually.\footnote{Note that this does not fully capture the systematic uncertainty associated with the Belle-II measurement, which includes other sources attributed to experimental effects that cannot be included in this study. Furthermore, the fit conducted here has larger statistical uncertainties due to the lack of access to full Monte Carlo samples. However, it should provide a reasonable estimate.} To validate these choices, a fit including only the SM contribution, and no injected NP signal, is first conducted. From this, Br$[B^+\to K^+\nu\bar{\nu}]=(2.8 \pm 0.7)\times10^{-5}$ is found, which is in good agreement with the result of the Belle~II analysis. 

The BaBar data are provided in bins of $S_B=q^2/m_B^2$ and the distributions are shown in Fig.~\ref{fig:babar_histograms} with contributions from the background, the SM to $B\to K\nu\nu$ and the NP signal. The associated signal efficiency in each bin is provided and considered in the fit via scaling of the resonance template. Only statistical uncertainties are accounted for in templates of the BaBar fit. In the case of the simultaneous fit to the Belle and BaBar data, the normalisation of the $B\to K\nu\nu$ templates are fixed by the SM expectations to the channels.

The NP signal is modelled with a Gaussian distribution, with the initial pre-fit yield of the template defined as the number of excess events observed in the Belle~II data. We assume a standard deviation of 1.5$\,$GeV$^2$ for this distribution, the result of this can be seen in Fig.~\ref{fig:signal_region_histograms}, where the NP contribution is scaled to the best-fit branching fraction. This assumption is also true of the fits to BaBar distributions, where the Gaussian mean and standard deviation are scaled as the other distributions ($S_B=q^2/m_B^2$).

Note that the look-elsewhere effect here is considered negligible since the fit only takes into account three (independent) $q^2$ bins (see left panel of Fig.~\ref{fig:signal_region_histograms}).

\acknowledgments{A.C.~gratefully acknowledges support by the Swiss National Science Foundation (No.\ PP00P21\_76884). G.I.~and H.H.~acknowledge support by the European Research Council under the grant agreement No. 947006 - \textit{InterLeptons}. The research of W.A.~is supported by the U.S.~Department of Energy grant number DE-SC0010107. J.M.C. thanks MICINN for funding through the grant ``DarkMaps'' PID2022-142142NB-I00. }
\newpage
\bibliography{bib}

\begin{thebibliography}{73}
\expandafter\ifx\csname natexlab\endcsname\relax\def\natexlab#1{#1}\fi
\expandafter\ifx\csname bibnamefont\endcsname\relax
  \def\bibnamefont#1{#1}\fi
\expandafter\ifx\csname bibfnamefont\endcsname\relax
  \def\bibfnamefont#1{#1}\fi
\expandafter\ifx\csname citenamefont\endcsname\relax
  \def\citenamefont#1{#1}\fi
\expandafter\ifx\csname url\endcsname\relax
  \def\url#1{\texttt{#1}}\fi
\expandafter\ifx\csname urlprefix\endcsname\relax\def\urlprefix{URL }\fi
\providecommand{\bibinfo}[2]{#2}
\providecommand{\eprint}[2][]{\url{#2}}

\bibitem[{\citenamefont{Kobayashi and Maskawa}(1973)}]{Kobayashi:1973fv}
\bibinfo{author}{\bibfnamefont{M.}~\bibnamefont{Kobayashi}} \bibnamefont{and}
  \bibinfo{author}{\bibfnamefont{T.}~\bibnamefont{Maskawa}},
  \bibinfo{journal}{Prog. Theor. Phys.} \textbf{\bibinfo{volume}{49}},
  \bibinfo{pages}{652} (\bibinfo{year}{1973}).

\bibitem[{\citenamefont{Abashian et~al.}(2002)}]{Belle:2000cnh}
\bibinfo{author}{\bibfnamefont{A.}~\bibnamefont{Abashian}} \bibnamefont{et~al.}
  (\bibinfo{collaboration}{Belle}), \bibinfo{journal}{Nucl. Instrum. Meth. A}
  \textbf{\bibinfo{volume}{479}}, \bibinfo{pages}{117} (\bibinfo{year}{2002}).

\bibitem[{\citenamefont{Aubert et~al.}(2002)}]{BaBar:2001yhh}
\bibinfo{author}{\bibfnamefont{B.}~\bibnamefont{Aubert}} \bibnamefont{et~al.}
  (\bibinfo{collaboration}{BaBar}), \bibinfo{journal}{Nucl. Instrum. Meth. A}
  \textbf{\bibinfo{volume}{479}}, \bibinfo{pages}{1} (\bibinfo{year}{2002}),
  \eprint{hep-ex/0105044}.

\bibitem[{\citenamefont{Higgs}(1964)}]{Higgs:1964ia}
\bibinfo{author}{\bibfnamefont{P.~W.} \bibnamefont{Higgs}},
  \bibinfo{journal}{Phys. Lett.} \textbf{\bibinfo{volume}{12}},
  \bibinfo{pages}{132} (\bibinfo{year}{1964}).

\bibitem[{\citenamefont{Englert and Brout}(1964)}]{Englert:1964et}
\bibinfo{author}{\bibfnamefont{F.}~\bibnamefont{Englert}} \bibnamefont{and}
  \bibinfo{author}{\bibfnamefont{R.}~\bibnamefont{Brout}},
  \bibinfo{journal}{Phys. Rev. Lett.} \textbf{\bibinfo{volume}{13}},
  \bibinfo{pages}{321} (\bibinfo{year}{1964}).

\bibitem[{\citenamefont{Aad et~al.}(2012)}]{ATLAS:2012yve}
\bibinfo{author}{\bibfnamefont{G.}~\bibnamefont{Aad}} \bibnamefont{et~al.}
  (\bibinfo{collaboration}{ATLAS}), \bibinfo{journal}{Phys. Lett. B}
  \textbf{\bibinfo{volume}{716}}, \bibinfo{pages}{1} (\bibinfo{year}{2012}),
  \eprint{1207.7214}.

\bibitem[{\citenamefont{Chatrchyan et~al.}(2012)}]{CMS:2012qbp}
\bibinfo{author}{\bibfnamefont{S.}~\bibnamefont{Chatrchyan}}
  \bibnamefont{et~al.} (\bibinfo{collaboration}{CMS}), \bibinfo{journal}{Phys.
  Lett. B} \textbf{\bibinfo{volume}{716}}, \bibinfo{pages}{30}
  (\bibinfo{year}{2012}), \eprint{1207.7235}.

\bibitem[{\citenamefont{Crivellin and Mellado}(2023)}]{Crivellin:2023zui}
\bibinfo{author}{\bibfnamefont{A.}~\bibnamefont{Crivellin}} \bibnamefont{and}
  \bibinfo{author}{\bibfnamefont{B.}~\bibnamefont{Mellado}}
  (\bibinfo{year}{2023}), \eprint{2309.03870}.

\bibitem[{\citenamefont{Albrecht et~al.}(2021)\citenamefont{Albrecht, van Dyk,
  and Langenbruch}}]{Albrecht:2021tul}
\bibinfo{author}{\bibfnamefont{J.}~\bibnamefont{Albrecht}},
  \bibinfo{author}{\bibfnamefont{D.}~\bibnamefont{van Dyk}}, \bibnamefont{and}
  \bibinfo{author}{\bibfnamefont{C.}~\bibnamefont{Langenbruch}},
  \bibinfo{journal}{Prog. Part. Nucl. Phys.} \textbf{\bibinfo{volume}{120}},
  \bibinfo{pages}{103885} (\bibinfo{year}{2021}), \eprint{2107.04822}.

\bibitem[{\citenamefont{London and Matias}(2022)}]{London:2021lfn}
\bibinfo{author}{\bibfnamefont{D.}~\bibnamefont{London}} \bibnamefont{and}
  \bibinfo{author}{\bibfnamefont{J.}~\bibnamefont{Matias}},
  \bibinfo{journal}{Ann. Rev. Nucl. Part. Sci.} \textbf{\bibinfo{volume}{72}},
  \bibinfo{pages}{37} (\bibinfo{year}{2022}), \eprint{2110.13270}.

\bibitem[{\citenamefont{Capdevila et~al.}(2023)\citenamefont{Capdevila,
  Crivellin, and Matias}}]{Capdevila:2023yhq}
\bibinfo{author}{\bibfnamefont{B.}~\bibnamefont{Capdevila}},
  \bibinfo{author}{\bibfnamefont{A.}~\bibnamefont{Crivellin}},
  \bibnamefont{and} \bibinfo{author}{\bibfnamefont{J.}~\bibnamefont{Matias}}
  (\bibinfo{year}{2023}), \eprint{2309.01311}.

\bibitem[{\citenamefont{Altmannshofer et~al.}(2009)\citenamefont{Altmannshofer,
  Buras, Straub, and Wick}}]{Altmannshofer:2009ma}
\bibinfo{author}{\bibfnamefont{W.}~\bibnamefont{Altmannshofer}},
  \bibinfo{author}{\bibfnamefont{A.~J.} \bibnamefont{Buras}},
  \bibinfo{author}{\bibfnamefont{D.~M.} \bibnamefont{Straub}},
  \bibnamefont{and} \bibinfo{author}{\bibfnamefont{M.}~\bibnamefont{Wick}},
  \bibinfo{journal}{JHEP} \textbf{\bibinfo{volume}{04}}, \bibinfo{pages}{022}
  (\bibinfo{year}{2009}), \eprint{0902.0160}.

\bibitem[{\citenamefont{Buras et~al.}(2015)\citenamefont{Buras, Girrbach-Noe,
  Niehoff, and Straub}}]{Buras:2014fpa}
\bibinfo{author}{\bibfnamefont{A.~J.} \bibnamefont{Buras}},
  \bibinfo{author}{\bibfnamefont{J.}~\bibnamefont{Girrbach-Noe}},
  \bibinfo{author}{\bibfnamefont{C.}~\bibnamefont{Niehoff}}, \bibnamefont{and}
  \bibinfo{author}{\bibfnamefont{D.~M.} \bibnamefont{Straub}},
  \bibinfo{journal}{JHEP} \textbf{\bibinfo{volume}{02}}, \bibinfo{pages}{184}
  (\bibinfo{year}{2015}), \eprint{1409.4557}.

\bibitem[{\citenamefont{Ganiev}(2023)}]{BptoKpnunuEPS}
\bibinfo{author}{\bibfnamefont{E.}~\bibnamefont{Ganiev}},
  \bibinfo{howpublished}{(On behalf of the Belle II Collaboration). Talk given
  at EPS-HEP 2023 “On Radiative and Electroweak Penguin Decays”, Hamburg,
  DESY
  (https://indico.desy.de/event/34916/contributions/-146877/attachments/84380/111798/-EWP@Belle2\_EPS.pdf)}
  (\bibinfo{year}{2023}).

\bibitem[{\citenamefont{Allwicher et~al.}(2023)\citenamefont{Allwicher,
  Becirevic, Piazza, Rosauro-Alcaraz, and Sumensari}}]{Allwicher:2023syp}
\bibinfo{author}{\bibfnamefont{L.}~\bibnamefont{Allwicher}},
  \bibinfo{author}{\bibfnamefont{D.}~\bibnamefont{Becirevic}},
  \bibinfo{author}{\bibfnamefont{G.}~\bibnamefont{Piazza}},
  \bibinfo{author}{\bibfnamefont{S.}~\bibnamefont{Rosauro-Alcaraz}},
  \bibnamefont{and} \bibinfo{author}{\bibfnamefont{O.}~\bibnamefont{Sumensari}}
  (\bibinfo{year}{2023}), \eprint{2309.02246}.

\bibitem[{\citenamefont{Athron et~al.}(2023)\citenamefont{Athron, Martinez, and
  Sierra}}]{Athron:2023hmz}
\bibinfo{author}{\bibfnamefont{P.}~\bibnamefont{Athron}},
  \bibinfo{author}{\bibfnamefont{R.}~\bibnamefont{Martinez}}, \bibnamefont{and}
  \bibinfo{author}{\bibfnamefont{C.}~\bibnamefont{Sierra}}
  (\bibinfo{year}{2023}), \eprint{2308.13426}.

\bibitem[{\citenamefont{Felkl et~al.}(2023)\citenamefont{Felkl, Giri, Mohanta,
  and Schmidt}}]{Felkl:2023ayn}
\bibinfo{author}{\bibfnamefont{T.}~\bibnamefont{Felkl}},
  \bibinfo{author}{\bibfnamefont{A.}~\bibnamefont{Giri}},
  \bibinfo{author}{\bibfnamefont{R.}~\bibnamefont{Mohanta}}, \bibnamefont{and}
  \bibinfo{author}{\bibfnamefont{M.~A.} \bibnamefont{Schmidt}}
  (\bibinfo{year}{2023}), \eprint{2309.02940}.

\bibitem[{\citenamefont{Chen and Chiang}(2023)}]{Chen:2023wpb}
\bibinfo{author}{\bibfnamefont{C.-H.} \bibnamefont{Chen}} \bibnamefont{and}
  \bibinfo{author}{\bibfnamefont{C.-W.} \bibnamefont{Chiang}}
  (\bibinfo{year}{2023}), \eprint{2309.12904}.

\bibitem[{\citenamefont{He et~al.}(2023)\citenamefont{He, Ma, and
  Valencia}}]{He:2023bnk}
\bibinfo{author}{\bibfnamefont{X.-G.} \bibnamefont{He}},
  \bibinfo{author}{\bibfnamefont{X.-D.} \bibnamefont{Ma}}, \bibnamefont{and}
  \bibinfo{author}{\bibfnamefont{G.}~\bibnamefont{Valencia}}
  (\bibinfo{year}{2023}), \eprint{2309.12741}.

\bibitem[{\citenamefont{Martin~Camalich
  et~al.}(2020)\citenamefont{Martin~Camalich, Pospelov, Vuong, Ziegler, and
  Zupan}}]{MartinCamalich:2020dfe}
\bibinfo{author}{\bibfnamefont{J.}~\bibnamefont{Martin~Camalich}},
  \bibinfo{author}{\bibfnamefont{M.}~\bibnamefont{Pospelov}},
  \bibinfo{author}{\bibfnamefont{P.~N.~H.} \bibnamefont{Vuong}},
  \bibinfo{author}{\bibfnamefont{R.}~\bibnamefont{Ziegler}}, \bibnamefont{and}
  \bibinfo{author}{\bibfnamefont{J.}~\bibnamefont{Zupan}},
  \bibinfo{journal}{Phys. Rev. D} \textbf{\bibinfo{volume}{102}},
  \bibinfo{pages}{015023} (\bibinfo{year}{2020}), \eprint{2002.04623}.

\bibitem[{\citenamefont{Guerrera and Rigolin}(2023)}]{Guerrera:2022ykl}
\bibinfo{author}{\bibfnamefont{A.~W.~M.} \bibnamefont{Guerrera}}
  \bibnamefont{and} \bibinfo{author}{\bibfnamefont{S.}~\bibnamefont{Rigolin}},
  \bibinfo{journal}{Fortsch. Phys.} \textbf{\bibinfo{volume}{71}},
  \bibinfo{pages}{2200192} (\bibinfo{year}{2023}), \eprint{2211.08343}.

\bibitem[{\citenamefont{Datta et~al.}(2023)\citenamefont{Datta, Hammad,
  Marfatia, Mukherjee, and Rashed}}]{Datta:2022zng}
\bibinfo{author}{\bibfnamefont{A.}~\bibnamefont{Datta}},
  \bibinfo{author}{\bibfnamefont{A.}~\bibnamefont{Hammad}},
  \bibinfo{author}{\bibfnamefont{D.}~\bibnamefont{Marfatia}},
  \bibinfo{author}{\bibfnamefont{L.}~\bibnamefont{Mukherjee}},
  \bibnamefont{and} \bibinfo{author}{\bibfnamefont{A.}~\bibnamefont{Rashed}},
  \bibinfo{journal}{JHEP} \textbf{\bibinfo{volume}{03}}, \bibinfo{pages}{108}
  (\bibinfo{year}{2023}), \eprint{2210.15662}.

\bibitem[{\citenamefont{Bruggisser et~al.}(2023)\citenamefont{Bruggisser,
  Grabitz, and Westhoff}}]{Bruggisser:2023npd}
\bibinfo{author}{\bibfnamefont{S.}~\bibnamefont{Bruggisser}},
  \bibinfo{author}{\bibfnamefont{L.}~\bibnamefont{Grabitz}}, \bibnamefont{and}
  \bibinfo{author}{\bibfnamefont{S.}~\bibnamefont{Westhoff}}
  (\bibinfo{year}{2023}), \eprint{2308.11703}.

\bibitem[{\citenamefont{Abdughani and Reyimuaji}(2023)}]{Abdughani:2023dlr}
\bibinfo{author}{\bibfnamefont{M.}~\bibnamefont{Abdughani}} \bibnamefont{and}
  \bibinfo{author}{\bibfnamefont{Y.}~\bibnamefont{Reyimuaji}}
  (\bibinfo{year}{2023}), \eprint{2309.03706}.

\bibitem[{\citenamefont{Berezhnoy and Melikhov}(2023)}]{Berezhnoy:2023rxx}
\bibinfo{author}{\bibfnamefont{A.}~\bibnamefont{Berezhnoy}} \bibnamefont{and}
  \bibinfo{author}{\bibfnamefont{D.}~\bibnamefont{Melikhov}}
  (\bibinfo{year}{2023}), \eprint{2309.17191}.

\bibitem[{\citenamefont{Kamenik and Smith}(2012)}]{Kamenik:2011vy}
\bibinfo{author}{\bibfnamefont{J.~F.} \bibnamefont{Kamenik}} \bibnamefont{and}
  \bibinfo{author}{\bibfnamefont{C.}~\bibnamefont{Smith}},
  \bibinfo{journal}{JHEP} \textbf{\bibinfo{volume}{03}}, \bibinfo{pages}{090}
  (\bibinfo{year}{2012}), \eprint{1111.6402}.

\bibitem[{\citenamefont{Goudzovski et~al.}(2023)}]{Goudzovski:2022vbt}
\bibinfo{author}{\bibfnamefont{E.}~\bibnamefont{Goudzovski}}
  \bibnamefont{et~al.}, \bibinfo{journal}{Rept. Prog. Phys.}
  \textbf{\bibinfo{volume}{86}}, \bibinfo{pages}{016201}
  (\bibinfo{year}{2023}), \eprint{2201.07805}.

\bibitem[{\citenamefont{Wilczek}(1982)}]{Wilczek:1982rv}
\bibinfo{author}{\bibfnamefont{F.}~\bibnamefont{Wilczek}},
  \bibinfo{journal}{Phys. Rev. Lett.} \textbf{\bibinfo{volume}{49}},
  \bibinfo{pages}{1549} (\bibinfo{year}{1982}).

\bibitem[{\citenamefont{Feng et~al.}(1998)\citenamefont{Feng, Moroi, Murayama,
  and Schnapka}}]{Feng:1997tn}
\bibinfo{author}{\bibfnamefont{J.~L.} \bibnamefont{Feng}},
  \bibinfo{author}{\bibfnamefont{T.}~\bibnamefont{Moroi}},
  \bibinfo{author}{\bibfnamefont{H.}~\bibnamefont{Murayama}}, \bibnamefont{and}
  \bibinfo{author}{\bibfnamefont{E.}~\bibnamefont{Schnapka}},
  \bibinfo{journal}{Phys. Rev. D} \textbf{\bibinfo{volume}{57}},
  \bibinfo{pages}{5875} (\bibinfo{year}{1998}), \eprint{hep-ph/9709411}.

\bibitem[{\citenamefont{Calibbi et~al.}(2017)\citenamefont{Calibbi, Goertz,
  Redigolo, Ziegler, and Zupan}}]{Calibbi:2016hwq}
\bibinfo{author}{\bibfnamefont{L.}~\bibnamefont{Calibbi}},
  \bibinfo{author}{\bibfnamefont{F.}~\bibnamefont{Goertz}},
  \bibinfo{author}{\bibfnamefont{D.}~\bibnamefont{Redigolo}},
  \bibinfo{author}{\bibfnamefont{R.}~\bibnamefont{Ziegler}}, \bibnamefont{and}
  \bibinfo{author}{\bibfnamefont{J.}~\bibnamefont{Zupan}},
  \bibinfo{journal}{Phys. Rev. D} \textbf{\bibinfo{volume}{95}},
  \bibinfo{pages}{095009} (\bibinfo{year}{2017}), \eprint{1612.08040}.

\bibitem[{\citenamefont{Ema et~al.}(2017)\citenamefont{Ema, Hamaguchi, Moroi,
  and Nakayama}}]{Ema:2016ops}
\bibinfo{author}{\bibfnamefont{Y.}~\bibnamefont{Ema}},
  \bibinfo{author}{\bibfnamefont{K.}~\bibnamefont{Hamaguchi}},
  \bibinfo{author}{\bibfnamefont{T.}~\bibnamefont{Moroi}}, \bibnamefont{and}
  \bibinfo{author}{\bibfnamefont{K.}~\bibnamefont{Nakayama}},
  \bibinfo{journal}{JHEP} \textbf{\bibinfo{volume}{01}}, \bibinfo{pages}{096}
  (\bibinfo{year}{2017}), \eprint{1612.05492}.

\bibitem[{\citenamefont{Arias-Aragon and Merlo}(2017)}]{Arias-Aragon:2017eww}
\bibinfo{author}{\bibfnamefont{F.}~\bibnamefont{Arias-Aragon}}
  \bibnamefont{and} \bibinfo{author}{\bibfnamefont{L.}~\bibnamefont{Merlo}},
  \bibinfo{journal}{JHEP} \textbf{\bibinfo{volume}{10}}, \bibinfo{pages}{168}
  (\bibinfo{year}{2017}), \bibinfo{note}{[Erratum: JHEP 11, 152 (2019)]},
  \eprint{1709.07039}.

\bibitem[{\citenamefont{Choi et~al.}(2017)\citenamefont{Choi, Im, Park, and
  Yun}}]{Choi:2017gpf}
\bibinfo{author}{\bibfnamefont{K.}~\bibnamefont{Choi}},
  \bibinfo{author}{\bibfnamefont{S.~H.} \bibnamefont{Im}},
  \bibinfo{author}{\bibfnamefont{C.~B.} \bibnamefont{Park}}, \bibnamefont{and}
  \bibinfo{author}{\bibfnamefont{S.}~\bibnamefont{Yun}},
  \bibinfo{journal}{JHEP} \textbf{\bibinfo{volume}{11}}, \bibinfo{pages}{070}
  (\bibinfo{year}{2017}), \eprint{1708.00021}.

\bibitem[{\citenamefont{Calibbi et~al.}(2021)\citenamefont{Calibbi, Redigolo,
  Ziegler, and Zupan}}]{Calibbi:2020jvd}
\bibinfo{author}{\bibfnamefont{L.}~\bibnamefont{Calibbi}},
  \bibinfo{author}{\bibfnamefont{D.}~\bibnamefont{Redigolo}},
  \bibinfo{author}{\bibfnamefont{R.}~\bibnamefont{Ziegler}}, \bibnamefont{and}
  \bibinfo{author}{\bibfnamefont{J.}~\bibnamefont{Zupan}},
  \bibinfo{journal}{JHEP} \textbf{\bibinfo{volume}{09}}, \bibinfo{pages}{173}
  (\bibinfo{year}{2021}), \eprint{2006.04795}.

\bibitem[{\citenamefont{Di~Luzio et~al.}(2023)\citenamefont{Di~Luzio, Guerrera,
  D\'\i{}az, and Rigolin}}]{DiLuzio:2023ndz}
\bibinfo{author}{\bibfnamefont{L.}~\bibnamefont{Di~Luzio}},
  \bibinfo{author}{\bibfnamefont{A.~W.~M.} \bibnamefont{Guerrera}},
  \bibinfo{author}{\bibfnamefont{X.~P.} \bibnamefont{D\'\i{}az}},
  \bibnamefont{and} \bibinfo{author}{\bibfnamefont{S.}~\bibnamefont{Rigolin}},
  \bibinfo{journal}{JHEP} \textbf{\bibinfo{volume}{06}}, \bibinfo{pages}{046}
  (\bibinfo{year}{2023}), \eprint{2304.04643}.

\bibitem[{\citenamefont{Gripaios et~al.}(2009)\citenamefont{Gripaios, Pomarol,
  Riva, and Serra}}]{Gripaios:2009pe}
\bibinfo{author}{\bibfnamefont{B.}~\bibnamefont{Gripaios}},
  \bibinfo{author}{\bibfnamefont{A.}~\bibnamefont{Pomarol}},
  \bibinfo{author}{\bibfnamefont{F.}~\bibnamefont{Riva}}, \bibnamefont{and}
  \bibinfo{author}{\bibfnamefont{J.}~\bibnamefont{Serra}},
  \bibinfo{journal}{JHEP} \textbf{\bibinfo{volume}{04}}, \bibinfo{pages}{070}
  (\bibinfo{year}{2009}), \eprint{0902.1483}.

\bibitem[{\citenamefont{Izaguirre et~al.}(2017)\citenamefont{Izaguirre, Lin,
  and Shuve}}]{Izaguirre:2016dfi}
\bibinfo{author}{\bibfnamefont{E.}~\bibnamefont{Izaguirre}},
  \bibinfo{author}{\bibfnamefont{T.}~\bibnamefont{Lin}}, \bibnamefont{and}
  \bibinfo{author}{\bibfnamefont{B.}~\bibnamefont{Shuve}},
  \bibinfo{journal}{Phys. Rev. Lett.} \textbf{\bibinfo{volume}{118}},
  \bibinfo{pages}{111802} (\bibinfo{year}{2017}), \eprint{1611.09355}.

\bibitem[{\citenamefont{Dolan et~al.}(2017)\citenamefont{Dolan, Ferber, Hearty,
  Kahlhoefer, and Schmidt-Hoberg}}]{Dolan:2017osp}
\bibinfo{author}{\bibfnamefont{M.~J.} \bibnamefont{Dolan}},
  \bibinfo{author}{\bibfnamefont{T.}~\bibnamefont{Ferber}},
  \bibinfo{author}{\bibfnamefont{C.}~\bibnamefont{Hearty}},
  \bibinfo{author}{\bibfnamefont{F.}~\bibnamefont{Kahlhoefer}},
  \bibnamefont{and}
  \bibinfo{author}{\bibfnamefont{K.}~\bibnamefont{Schmidt-Hoberg}},
  \bibinfo{journal}{JHEP} \textbf{\bibinfo{volume}{12}}, \bibinfo{pages}{094}
  (\bibinfo{year}{2017}), \bibinfo{note}{[Erratum: JHEP 03, 190 (2021)]},
  \eprint{1709.00009}.

\bibitem[{\citenamefont{Bj\"orkeroth et~al.}(2018)\citenamefont{Bj\"orkeroth,
  Chun, and King}}]{Bjorkeroth:2018dzu}
\bibinfo{author}{\bibfnamefont{F.}~\bibnamefont{Bj\"orkeroth}},
  \bibinfo{author}{\bibfnamefont{E.~J.} \bibnamefont{Chun}}, \bibnamefont{and}
  \bibinfo{author}{\bibfnamefont{S.~F.} \bibnamefont{King}},
  \bibinfo{journal}{JHEP} \textbf{\bibinfo{volume}{08}}, \bibinfo{pages}{117}
  (\bibinfo{year}{2018}), \eprint{1806.00660}.

\bibitem[{\citenamefont{Gavela et~al.}(2019)\citenamefont{Gavela, Houtz,
  Quilez, Del~Rey, and Sumensari}}]{Gavela:2019wzg}
\bibinfo{author}{\bibfnamefont{M.~B.} \bibnamefont{Gavela}},
  \bibinfo{author}{\bibfnamefont{R.}~\bibnamefont{Houtz}},
  \bibinfo{author}{\bibfnamefont{P.}~\bibnamefont{Quilez}},
  \bibinfo{author}{\bibfnamefont{R.}~\bibnamefont{Del~Rey}}, \bibnamefont{and}
  \bibinfo{author}{\bibfnamefont{O.}~\bibnamefont{Sumensari}},
  \bibinfo{journal}{Eur. Phys. J. C} \textbf{\bibinfo{volume}{79}},
  \bibinfo{pages}{369} (\bibinfo{year}{2019}), \eprint{1901.02031}.

\bibitem[{\citenamefont{Carmona et~al.}(2021)\citenamefont{Carmona, Scherb, and
  Schwaller}}]{Carmona:2021seb}
\bibinfo{author}{\bibfnamefont{A.}~\bibnamefont{Carmona}},
  \bibinfo{author}{\bibfnamefont{C.}~\bibnamefont{Scherb}}, \bibnamefont{and}
  \bibinfo{author}{\bibfnamefont{P.}~\bibnamefont{Schwaller}},
  \bibinfo{journal}{JHEP} \textbf{\bibinfo{volume}{08}}, \bibinfo{pages}{121}
  (\bibinfo{year}{2021}), \eprint{2101.07803}.

\bibitem[{\citenamefont{Bauer et~al.}(2022)\citenamefont{Bauer, Neubert,
  Renner, Schnubel, and Thamm}}]{Bauer:2021mvw}
\bibinfo{author}{\bibfnamefont{M.}~\bibnamefont{Bauer}},
  \bibinfo{author}{\bibfnamefont{M.}~\bibnamefont{Neubert}},
  \bibinfo{author}{\bibfnamefont{S.}~\bibnamefont{Renner}},
  \bibinfo{author}{\bibfnamefont{M.}~\bibnamefont{Schnubel}}, \bibnamefont{and}
  \bibinfo{author}{\bibfnamefont{A.}~\bibnamefont{Thamm}},
  \bibinfo{journal}{JHEP} \textbf{\bibinfo{volume}{09}}, \bibinfo{pages}{056}
  (\bibinfo{year}{2022}), \eprint{2110.10698}.

\bibitem[{\citenamefont{Ferber et~al.}(2023)\citenamefont{Ferber, Filimonova,
  Sch\"afer, and Westhoff}}]{Ferber:2022rsf}
\bibinfo{author}{\bibfnamefont{T.}~\bibnamefont{Ferber}},
  \bibinfo{author}{\bibfnamefont{A.}~\bibnamefont{Filimonova}},
  \bibinfo{author}{\bibfnamefont{R.}~\bibnamefont{Sch\"afer}},
  \bibnamefont{and} \bibinfo{author}{\bibfnamefont{S.}~\bibnamefont{Westhoff}},
  \bibinfo{journal}{JHEP} \textbf{\bibinfo{volume}{04}}, \bibinfo{pages}{131}
  (\bibinfo{year}{2023}), \eprint{2201.06580}.

\bibitem[{\citenamefont{Sala and Straub}(2017)}]{Sala:2017ihs}
\bibinfo{author}{\bibfnamefont{F.}~\bibnamefont{Sala}} \bibnamefont{and}
  \bibinfo{author}{\bibfnamefont{D.~M.} \bibnamefont{Straub}},
  \bibinfo{journal}{Phys. Lett. B} \textbf{\bibinfo{volume}{774}},
  \bibinfo{pages}{205} (\bibinfo{year}{2017}), \eprint{1704.06188}.

\bibitem[{\citenamefont{Mohapatra and Giri}(2021)}]{Mohapatra:2021izl}
\bibinfo{author}{\bibfnamefont{M.~K.} \bibnamefont{Mohapatra}}
  \bibnamefont{and} \bibinfo{author}{\bibfnamefont{A.}~\bibnamefont{Giri}},
  \bibinfo{journal}{Phys. Rev. D} \textbf{\bibinfo{volume}{104}},
  \bibinfo{pages}{095012} (\bibinfo{year}{2021}), \eprint{2109.12382}.

\bibitem[{\citenamefont{Datta et~al.}(2018)\citenamefont{Datta, Kumar, Liao,
  and Marfatia}}]{Datta:2017ezo}
\bibinfo{author}{\bibfnamefont{A.}~\bibnamefont{Datta}},
  \bibinfo{author}{\bibfnamefont{J.}~\bibnamefont{Kumar}},
  \bibinfo{author}{\bibfnamefont{J.}~\bibnamefont{Liao}}, \bibnamefont{and}
  \bibinfo{author}{\bibfnamefont{D.}~\bibnamefont{Marfatia}},
  \bibinfo{journal}{Phys. Rev. D} \textbf{\bibinfo{volume}{97}},
  \bibinfo{pages}{115038} (\bibinfo{year}{2018}), \eprint{1705.08423}.

\bibitem[{\citenamefont{Altmannshofer et~al.}(2018)\citenamefont{Altmannshofer,
  Baker, Gori, Harnik, Pospelov, Stamou, and Thamm}}]{Altmannshofer:2017bsz}
\bibinfo{author}{\bibfnamefont{W.}~\bibnamefont{Altmannshofer}},
  \bibinfo{author}{\bibfnamefont{M.~J.} \bibnamefont{Baker}},
  \bibinfo{author}{\bibfnamefont{S.}~\bibnamefont{Gori}},
  \bibinfo{author}{\bibfnamefont{R.}~\bibnamefont{Harnik}},
  \bibinfo{author}{\bibfnamefont{M.}~\bibnamefont{Pospelov}},
  \bibinfo{author}{\bibfnamefont{E.}~\bibnamefont{Stamou}}, \bibnamefont{and}
  \bibinfo{author}{\bibfnamefont{A.}~\bibnamefont{Thamm}},
  \bibinfo{journal}{JHEP} \textbf{\bibinfo{volume}{03}}, \bibinfo{pages}{188}
  (\bibinfo{year}{2018}), \eprint{1711.07494}.

\bibitem[{\citenamefont{Sala}(2018)}]{Sala:2018ukk}
\bibinfo{author}{\bibfnamefont{F.}~\bibnamefont{Sala}}, \bibinfo{journal}{Nucl.
  Part. Phys. Proc.} \textbf{\bibinfo{volume}{303-305}}, \bibinfo{pages}{14}
  (\bibinfo{year}{2018}), \eprint{1809.11061}.

\bibitem[{\citenamefont{Bishara et~al.}(2017)\citenamefont{Bishara, Haisch, and
  Monni}}]{Bishara:2017pje}
\bibinfo{author}{\bibfnamefont{F.}~\bibnamefont{Bishara}},
  \bibinfo{author}{\bibfnamefont{U.}~\bibnamefont{Haisch}}, \bibnamefont{and}
  \bibinfo{author}{\bibfnamefont{P.~F.} \bibnamefont{Monni}},
  \bibinfo{journal}{Phys. Rev. D} \textbf{\bibinfo{volume}{96}},
  \bibinfo{pages}{055002} (\bibinfo{year}{2017}), \eprint{1705.03465}.

\bibitem[{\citenamefont{Borah et~al.}(2020)\citenamefont{Borah, Mukherjee, and
  Nandi}}]{Borah:2020swo}
\bibinfo{author}{\bibfnamefont{D.}~\bibnamefont{Borah}},
  \bibinfo{author}{\bibfnamefont{L.}~\bibnamefont{Mukherjee}},
  \bibnamefont{and} \bibinfo{author}{\bibfnamefont{S.}~\bibnamefont{Nandi}},
  \bibinfo{journal}{JHEP} \textbf{\bibinfo{volume}{12}}, \bibinfo{pages}{052}
  (\bibinfo{year}{2020}), \eprint{2007.13778}.

\bibitem[{\citenamefont{Darm\'e et~al.}(2022)\citenamefont{Darm\'e, Fedele,
  Kowalska, and Sessolo}}]{Darme:2021qzw}
\bibinfo{author}{\bibfnamefont{L.}~\bibnamefont{Darm\'e}},
  \bibinfo{author}{\bibfnamefont{M.}~\bibnamefont{Fedele}},
  \bibinfo{author}{\bibfnamefont{K.}~\bibnamefont{Kowalska}}, \bibnamefont{and}
  \bibinfo{author}{\bibfnamefont{E.~M.} \bibnamefont{Sessolo}},
  \bibinfo{journal}{JHEP} \textbf{\bibinfo{volume}{03}}, \bibinfo{pages}{085}
  (\bibinfo{year}{2022}), \eprint{2106.12582}.

\bibitem[{\citenamefont{Greljo et~al.}(2022)\citenamefont{Greljo, Soreq,
  Stangl, Thomsen, and Zupan}}]{Greljo:2021npi}
\bibinfo{author}{\bibfnamefont{A.}~\bibnamefont{Greljo}},
  \bibinfo{author}{\bibfnamefont{Y.}~\bibnamefont{Soreq}},
  \bibinfo{author}{\bibfnamefont{P.}~\bibnamefont{Stangl}},
  \bibinfo{author}{\bibfnamefont{A.~E.} \bibnamefont{Thomsen}},
  \bibnamefont{and} \bibinfo{author}{\bibfnamefont{J.}~\bibnamefont{Zupan}},
  \bibinfo{journal}{JHEP} \textbf{\bibinfo{volume}{04}}, \bibinfo{pages}{151}
  (\bibinfo{year}{2022}), \eprint{2107.07518}.

\bibitem[{\citenamefont{Crivellin et~al.}(2022)\citenamefont{Crivellin,
  Manzari, Altmannshofer, Inguglia, Feichtinger, and
  Martin~Camalich}}]{Crivellin:2022obd}
\bibinfo{author}{\bibfnamefont{A.}~\bibnamefont{Crivellin}},
  \bibinfo{author}{\bibfnamefont{C.~A.} \bibnamefont{Manzari}},
  \bibinfo{author}{\bibfnamefont{W.}~\bibnamefont{Altmannshofer}},
  \bibinfo{author}{\bibfnamefont{G.}~\bibnamefont{Inguglia}},
  \bibinfo{author}{\bibfnamefont{P.}~\bibnamefont{Feichtinger}},
  \bibnamefont{and}
  \bibinfo{author}{\bibfnamefont{J.}~\bibnamefont{Martin~Camalich}},
  \bibinfo{journal}{Phys. Rev. D} \textbf{\bibinfo{volume}{106}},
  \bibinfo{pages}{L031703} (\bibinfo{year}{2022}), \eprint{2202.12900}.

\bibitem[{\citenamefont{Lees et~al.}(2013)}]{BaBar:2013npw}
\bibinfo{author}{\bibfnamefont{J.~P.} \bibnamefont{Lees}} \bibnamefont{et~al.}
  (\bibinfo{collaboration}{BaBar}), \bibinfo{journal}{Phys. Rev. D}
  \textbf{\bibinfo{volume}{87}}, \bibinfo{pages}{112005}
  (\bibinfo{year}{2013}), \eprint{1303.7465}.

\bibitem[{\citenamefont{del Amo~Sanchez et~al.}(2010)}]{BaBar:2010oqg}
\bibinfo{author}{\bibfnamefont{P.}~\bibnamefont{del Amo~Sanchez}}
  \bibnamefont{et~al.} (\bibinfo{collaboration}{BaBar}),
  \bibinfo{journal}{Phys. Rev. D} \textbf{\bibinfo{volume}{82}},
  \bibinfo{pages}{112002} (\bibinfo{year}{2010}), \eprint{1009.1529}.

\bibitem[{\citenamefont{Lutz et~al.}(2013)}]{Belle:2013tnz}
\bibinfo{author}{\bibfnamefont{O.}~\bibnamefont{Lutz}} \bibnamefont{et~al.}
  (\bibinfo{collaboration}{Belle}), \bibinfo{journal}{Phys. Rev. D}
  \textbf{\bibinfo{volume}{87}}, \bibinfo{pages}{111103}
  (\bibinfo{year}{2013}), \eprint{1303.3719}.

\bibitem[{\citenamefont{Grygier et~al.}(2017)}]{Belle:2017oht}
\bibinfo{author}{\bibfnamefont{J.}~\bibnamefont{Grygier}} \bibnamefont{et~al.}
  (\bibinfo{collaboration}{Belle}), \bibinfo{journal}{Phys. Rev. D}
  \textbf{\bibinfo{volume}{96}}, \bibinfo{pages}{091101}
  (\bibinfo{year}{2017}), \bibinfo{note}{[Addendum: Phys.Rev.D 97, 099902
  (2018)]}, \eprint{1702.03224}.

\bibitem[{\citenamefont{Prim et~al.}(2020)}]{Belle:2019iji}
\bibinfo{author}{\bibfnamefont{M.~T.} \bibnamefont{Prim}} \bibnamefont{et~al.}
  (\bibinfo{collaboration}{Belle}), \bibinfo{journal}{Phys. Rev. D}
  \textbf{\bibinfo{volume}{101}}, \bibinfo{pages}{032007}
  (\bibinfo{year}{2020}), \eprint{1911.03186}.

\bibitem[{\citenamefont{Abudin\'en et~al.}(2021)}]{Belle-II:2021rof}
\bibinfo{author}{\bibfnamefont{F.}~\bibnamefont{Abudin\'en}}
  \bibnamefont{et~al.} (\bibinfo{collaboration}{Belle-II}),
  \bibinfo{journal}{Phys. Rev. Lett.} \textbf{\bibinfo{volume}{127}},
  \bibinfo{pages}{181802} (\bibinfo{year}{2021}), \eprint{2104.12624}.

\bibitem[{\citenamefont{Parrott
  et~al.}(2023{\natexlab{a}})\citenamefont{Parrott, Bouchard, and
  Davies}}]{Parrott:2022zte}
\bibinfo{author}{\bibfnamefont{W.~G.} \bibnamefont{Parrott}},
  \bibinfo{author}{\bibfnamefont{C.}~\bibnamefont{Bouchard}}, \bibnamefont{and}
  \bibinfo{author}{\bibfnamefont{C.~T.~H.} \bibnamefont{Davies}}
  (\bibinfo{collaboration}{HPQCD}), \bibinfo{journal}{Phys. Rev. D}
  \textbf{\bibinfo{volume}{107}}, \bibinfo{pages}{014511}
  (\bibinfo{year}{2023}{\natexlab{a}}), \bibinfo{note}{[Erratum: Phys.Rev.D
  107, 119903 (2023)]}, \eprint{2207.13371}.

\bibitem[{\citenamefont{Be\v{c}irevi\'c
  et~al.}(2023)\citenamefont{Be\v{c}irevi\'c, Piazza, and
  Sumensari}}]{Becirevic:2023aov}
\bibinfo{author}{\bibfnamefont{D.}~\bibnamefont{Be\v{c}irevi\'c}},
  \bibinfo{author}{\bibfnamefont{G.}~\bibnamefont{Piazza}}, \bibnamefont{and}
  \bibinfo{author}{\bibfnamefont{O.}~\bibnamefont{Sumensari}},
  \bibinfo{journal}{Eur. Phys. J. C} \textbf{\bibinfo{volume}{83}},
  \bibinfo{pages}{252} (\bibinfo{year}{2023}), \eprint{2301.06990}.

\bibitem[{\citenamefont{Amhis et~al.}(2023)\citenamefont{Amhis, Kenzie, Reboud,
  and Wiederhold}}]{Amhis:2023mpj}
\bibinfo{author}{\bibfnamefont{Y.}~\bibnamefont{Amhis}},
  \bibinfo{author}{\bibfnamefont{M.}~\bibnamefont{Kenzie}},
  \bibinfo{author}{\bibfnamefont{M.}~\bibnamefont{Reboud}}, \bibnamefont{and}
  \bibinfo{author}{\bibfnamefont{A.~R.} \bibnamefont{Wiederhold}}
  (\bibinfo{year}{2023}), \eprint{2309.11353}.

\bibitem[{\citenamefont{Heinrich et~al.}(2021)\citenamefont{Heinrich, Feickert,
  Stark, and Cranmer}}]{Heinrich:2021gyp}
\bibinfo{author}{\bibfnamefont{L.}~\bibnamefont{Heinrich}},
  \bibinfo{author}{\bibfnamefont{M.}~\bibnamefont{Feickert}},
  \bibinfo{author}{\bibfnamefont{G.}~\bibnamefont{Stark}}, \bibnamefont{and}
  \bibinfo{author}{\bibfnamefont{K.}~\bibnamefont{Cranmer}},
  \bibinfo{journal}{J. Open Source Softw.} \textbf{\bibinfo{volume}{6}},
  \bibinfo{pages}{2823} (\bibinfo{year}{2021}).

\bibitem[{\citenamefont{Parrott
  et~al.}(2023{\natexlab{b}})\citenamefont{Parrott, Bouchard, and
  Davies}}]{Parrott:2022rgu}
\bibinfo{author}{\bibfnamefont{W.~G.} \bibnamefont{Parrott}},
  \bibinfo{author}{\bibfnamefont{C.}~\bibnamefont{Bouchard}}, \bibnamefont{and}
  \bibinfo{author}{\bibfnamefont{C.~T.~H.} \bibnamefont{Davies}}
  (\bibinfo{collaboration}{(HPQCD collaboration)\textsection{}, HPQCD}),
  \bibinfo{journal}{Phys. Rev. D} \textbf{\bibinfo{volume}{107}},
  \bibinfo{pages}{014510} (\bibinfo{year}{2023}{\natexlab{b}}),
  \eprint{2207.12468}.

\bibitem[{\citenamefont{Gubernari et~al.}(2023)\citenamefont{Gubernari, Reboud,
  van Dyk, and Virto}}]{Gubernari:2023puw}
\bibinfo{author}{\bibfnamefont{N.}~\bibnamefont{Gubernari}},
  \bibinfo{author}{\bibfnamefont{M.}~\bibnamefont{Reboud}},
  \bibinfo{author}{\bibfnamefont{D.}~\bibnamefont{van Dyk}}, \bibnamefont{and}
  \bibinfo{author}{\bibfnamefont{J.}~\bibnamefont{Virto}}
  (\bibinfo{year}{2023}), \eprint{2305.06301}.

\bibitem[{\citenamefont{Bharucha et~al.}(2016)\citenamefont{Bharucha, Straub,
  and Zwicky}}]{Bharucha:2015bzk}
\bibinfo{author}{\bibfnamefont{A.}~\bibnamefont{Bharucha}},
  \bibinfo{author}{\bibfnamefont{D.~M.} \bibnamefont{Straub}},
  \bibnamefont{and} \bibinfo{author}{\bibfnamefont{R.}~\bibnamefont{Zwicky}},
  \bibinfo{journal}{JHEP} \textbf{\bibinfo{volume}{08}}, \bibinfo{pages}{098}
  (\bibinfo{year}{2016}), \eprint{1503.05534}.

\bibitem[{\citenamefont{Altmannshofer et~al.}(2014)\citenamefont{Altmannshofer,
  Gori, Pospelov, and Yavin}}]{Altmannshofer:2014cfa}
\bibinfo{author}{\bibfnamefont{W.}~\bibnamefont{Altmannshofer}},
  \bibinfo{author}{\bibfnamefont{S.}~\bibnamefont{Gori}},
  \bibinfo{author}{\bibfnamefont{M.}~\bibnamefont{Pospelov}}, \bibnamefont{and}
  \bibinfo{author}{\bibfnamefont{I.}~\bibnamefont{Yavin}},
  \bibinfo{journal}{Phys. Rev. D} \textbf{\bibinfo{volume}{89}},
  \bibinfo{pages}{095033} (\bibinfo{year}{2014}), \eprint{1403.1269}.

\bibitem[{\citenamefont{Crivellin et~al.}(2015)\citenamefont{Crivellin,
  D'Ambrosio, and Heeck}}]{Crivellin:2015lwa}
\bibinfo{author}{\bibfnamefont{A.}~\bibnamefont{Crivellin}},
  \bibinfo{author}{\bibfnamefont{G.}~\bibnamefont{D'Ambrosio}},
  \bibnamefont{and} \bibinfo{author}{\bibfnamefont{J.}~\bibnamefont{Heeck}},
  \bibinfo{journal}{Phys. Rev. D} \textbf{\bibinfo{volume}{91}},
  \bibinfo{pages}{075006} (\bibinfo{year}{2015}), \eprint{1503.03477}.

\bibitem[{\citenamefont{Dreiner et~al.}(2023)\citenamefont{Dreiner, G\"unther,
  and Wang}}]{Dreiner:2023cms}
\bibinfo{author}{\bibfnamefont{H.~K.} \bibnamefont{Dreiner}},
  \bibinfo{author}{\bibfnamefont{J.~Y.} \bibnamefont{G\"unther}},
  \bibnamefont{and} \bibinfo{author}{\bibfnamefont{Z.~S.} \bibnamefont{Wang}}
  (\bibinfo{year}{2023}), \eprint{2309.03727}.

\bibitem[{\citenamefont{Adachi et~al.}(2020)}]{Adachi:2019otg}
\bibinfo{author}{\bibfnamefont{I.}~\bibnamefont{Adachi}} \bibnamefont{et~al.}
  (\bibinfo{collaboration}{Belle-II}), \bibinfo{journal}{Phys. Rev. Lett.}
  \textbf{\bibinfo{volume}{124}}, \bibinfo{pages}{141801}
  (\bibinfo{year}{2020}), \eprint{1912.11276}.

\bibitem[{\citenamefont{Browder et~al.}(2001)}]{CLEO:2000yzg}
\bibinfo{author}{\bibfnamefont{T.~E.} \bibnamefont{Browder}}
  \bibnamefont{et~al.} (\bibinfo{collaboration}{CLEO}), \bibinfo{journal}{Phys.
  Rev. Lett.} \textbf{\bibinfo{volume}{86}}, \bibinfo{pages}{2950}
  (\bibinfo{year}{2001}), \eprint{hep-ex/0007057}.

\bibitem[{\citenamefont{He et~al.}(1991)\citenamefont{He, Joshi, Lew, and
  Volkas}}]{He:1991qd}
\bibinfo{author}{\bibfnamefont{X.-G.} \bibnamefont{He}},
  \bibinfo{author}{\bibfnamefont{G.~C.} \bibnamefont{Joshi}},
  \bibinfo{author}{\bibfnamefont{H.}~\bibnamefont{Lew}}, \bibnamefont{and}
  \bibinfo{author}{\bibfnamefont{R.~R.} \bibnamefont{Volkas}},
  \bibinfo{journal}{Phys. Rev. D} \textbf{\bibinfo{volume}{44}},
  \bibinfo{pages}{2118} (\bibinfo{year}{1991}).

\bibitem[{\citenamefont{Crivellin et~al.}(2013)\citenamefont{Crivellin, Kokulu,
  and Greub}}]{Crivellin:2013wna}
\bibinfo{author}{\bibfnamefont{A.}~\bibnamefont{Crivellin}},
  \bibinfo{author}{\bibfnamefont{A.}~\bibnamefont{Kokulu}}, \bibnamefont{and}
  \bibinfo{author}{\bibfnamefont{C.}~\bibnamefont{Greub}},
  \bibinfo{journal}{Phys. Rev. D} \textbf{\bibinfo{volume}{87}},
  \bibinfo{pages}{094031} (\bibinfo{year}{2013}), \eprint{1303.5877}.

\end{thebibliography}
\end{document}